\documentclass[preprint,article,submit,pdftex,moreauthors]{mdpi} 
\firstpage{1} 
\pubvolume{1}
\issuenum{1}
\articlenumber{0}
\pubyear{2022}
\copyrightyear{2022}
\datereceived{} 
\dateaccepted{} 
\datepublished{} 
\hreflink{https://doi.org/} 

\newcommand\FS{\Phi} 
\newcommand\FSa{\Psi} 
\newcommand\parent{{\uparrow}} 
\newcommand\setm{\smallsetminus}
\newcommand\N{\mathbb N} 
\newcommand\E{\mathbb O} 
\newcommand\F{\mathbb F} 
\newcommand\D{\mathbb D} 
\newcommand\f{\mathsf f} 
\newcommand\orel{\ll} 
\newcommand\indep{\mathbin{\mkern-.3\thinmuskip\parallel\mkern-.3\thinmuskip}}
\newcommand\notindep{\mathbin{\mkern-.3\thinmuskip\nparallel\mkern-.3\thinmuskip}}
\newcommand\subc{\Subset} 
\newcommand\no{\mathsf{ok}}\newcommand\yes{\mathsf{conf}}
\let\eps\epsilon  
\let\preceq\preccurlyeq  
\DeclareFontFamily{U}{mathb}{\hyphenchar\font45}
\DeclareFontShape{U}{mathb}{m}{n}{
      <5> <6> <7> <8> <9> <10> gen * mathb
      <10.95> mathb10 <12> <14.4> <17.28> <20.74> <24.88> mathb12
      }{}
\DeclareSymbolFont{mathb}{U}{mathb}{m}{n}
\DeclareFontSubstitution{U}{mathb}{m}{n}
\DeclareMathSymbol\semle{3}{mathb}{"84}  
\DeclareMathSymbol\semge{3}{mathb}{"85}

\newcommand\undersc{\_}
\def\(#1,#2,#3){\langle #1,#2,#3\rangle} 
\ifthenelse{\isundefined\linenomath}{\let\linenomath\relax\let\endlinenomath\relax}{}
\newenvironment{disp}{\linenomath$$}{$$\endlinenomath\ignorespaces}
\newenvironment{itemz}[1][3pt]{%
\setitemize{topsep=#1,noitemsep,leftmargin=\parindent,labelwidth=\parindent,labelsep=0pt,align=parleft}%
\begin{itemize}}{\end{itemize}}

\Title{Data Synchronization: A Complete Theoretical Solution for Filesystems}

\TitleCitation{Data Synchronization: A Complete Theoretical Solution for Filesystems}

\Author{Elod P. Csirmaz$^{1,*}$\orcidA{} and Laszlo Csirmaz$^{2}$\orcidB{}}

\AuthorNames{Elod P. Csirmaz and Laszlo Csirmaz}

\AuthorCitation{Csirmaz, E. P; Csirmaz, L}

\address{%
$^{1}$ \quad elod@epcsirmaz.com\\
$^{2}$ \quad R\'enyi Institute, Budapest, and UTIA, Prague; csirmaz@renyi.hu}

\corres{Correspondence: elod@epcsirmaz.com}

\abstract{%
\fontsize{9}{10.8}\selectfont
Data reconciliation in general, and filesystem synchronization in
particular, lacks rigorous theoretical foundation. This paper presents, for
the first time, a complete analysis of synchronization for two replicas of a
theoretical filesystem. Synchronization has two main stages: identifying the
conflicts, and resolving them. All existing (both theoretical and practical)
synchronizers are \emph{operation-based:} they define, using some rationale
or heuristics, how conflicts are to be resolved without considering the
effect of the resolution on subsequent conflicts. Instead, our approach is
\emph{declaration-based:} we define what constitutes the resolution of all
conflicts, and for each possible scenario we prove the existence of
sequences of operations/commands which convert the replicas into a common
synchronized state. These sequences consist of operations rolling back some
local changes, followed by operations performed on the other replica. The
set of rolled-back operations provides the user with clear and intuitive
information on the proposed changes, so she can easily decide whether to
accept them or ask for other alternatives. All possible synchronized states
are described by specifying a set of conflicts, a partial order on the
conflicts describing the order in which they need to be resolved, as well as
the effect of each decision on subsequent conflicts. Using this
classification, the outcomes of different conflict resolution policies can
be investigated easily.
}



\begin{document}
\def\MM{1.1} 
\newcommand\Dfive[5]{
\foreach\x in{0,...,4}{
   \draw[color=black!50!white] (\x*\MM,0) circle(0.3);
}
\foreach\x in{0,...,3}{
   \draw[<-,thick] (\x*\MM+0.3,0) -- ({(\x+1)*\MM-0.3},0);
}
\draw (0,0) node {$#1$}
      (\MM,0) node {$#2$}
      (2*\MM,0) node {$#3$}
      (3*\MM,0) node {$#4$}
      (4*\MM,0) node {$#5$};
}
\newcommand\Dnine[9]{
\Dfive{#1}{#2}{#3}{#4}{#5}
\foreach \x in {1,...,5}{
  \draw[color=black!60!white] (\x*\MM+0.28-\MM,0.35) node {$\scriptstyle n_\x$};
}
\foreach\y in {-\MM}{
\foreach\x in {0,1,2,3}{
   \draw[color=black!50!white] (\x*\MM,\y) circle(0.3);
}
\foreach \x in {6,7,8,9}{
  \draw[color=black!60!white] (\x*\MM+0.28-6*\MM,\y+0.35) node {$\scriptstyle n_\x$};
}
\foreach\x in {0,1,2,3}{
   \draw[<-,thick] (\x*\MM,-0.3) -- (\x*\MM,\y+0.3);
}
\draw (0,\y) node {$#6$}
      (\MM,\y) node {$#7$}
      (2*\MM,\y) node {$#8$}
      (3*\MM,\y) node {$#9$};
}}
\newcommand\RAB[1]{
\draw[color=red!50!black,line width=1pt,<-]
  (#1*\MM-\MM+0.3,0.2) to[out=45,in=135] (#1*\MM-0.3,0.2);
}
\newcommand\RAF[1]{
\draw[color=red!50!black,line width=1pt,->]
  (#1*\MM+0.3,0.2) to[out=45,in=135] (#1*\MM+\MM-0.3,0.2);
}
\newcommand\DBOX[1]{ 
\draw[color=black!20!white,dotted,line width=1.4pt]
   (#1*\MM-0.4,-0.2 ) -- (#1*\MM-0.4,1.55) to[out=90,in=90] (#1*\MM+0.4,1.55)
   -- (#1*\MM+0.4,-0.2) to[out=270,in=270] (#1*\MM-0.4,-0.2);
}

\section{Introduction and related work}\label{sec:intro}

This work is a comprehensive treatment of filesystem synchronization and
conflict resolution on a simple but powerful theoretical model of
filesystems. Synchronizing diverged copies of some data stored on a variety
of devices and locations is an important and ubiquitous task. The last two
decades have seen tremendous advancement, both theoretical and practical, in
the closely related fields of distributed data storage
\cite{CRDT-overview,CRDT-orig} and group editors \cite{SJZ98,SE98}. This
progress has been based on, and has expanded significantly, two competing
theoretical frameworks: Operational Transformation (OT) and Conflict-free
Replicated Data Types (CRDT). OT appeared in the seminal work \cite{EG89}
and was refined later in \cite{SE98}. The general idea is that operations
are enriched with the context in which they were generated. Before applying
them, they are \emph{transformed} depending on the context of their origin
and the context where they are to be applied. Main applications are
collaborative editors, the most notable example being Google Docs
\cite{DR10}. The core concept of CRDT is \emph{commutativity}, see
\cite{CRDT-orig,CRDT-overview}. Basic data types with special operators are
devised so that executing the operators in different orders yield the same
results. Examples are counters or sets with add and delete operators. The
basic types are used as building blocks in more complex applications such as
collaborative editors \cite{NIC16}, the commercial product Riak
\cite{KLO10}, and others.

\begin{figure}[H]\centering
\begin{tikzpicture}[shorten >=2pt,scale=0.98]
\def\ss{\scriptstyle}
\node[draw] (q0) at (-1.5,0) {$\FS$};
\node[draw] (q1) at (0,2.2) {$\FS$};
\node[draw] (q2) at (0,-2.2) {$\FS$};
\node[draw] (q3) at (3.5,2.2) {$\FS_1$};
\node[draw] (q4) at (3.5,-2.2) {$\FS_2$};
\node[draw] (q5) at (7.2,2.2) {$\FSa$};
\node[draw] (q6) at (7.2,-2.2) {$\FSa$};
\node[draw] (q7) at (9,0) {$\FSa$};
\draw (7.2,0) node[rotate=90]{\footnotesize\dots\dots~~merged state~~\dots\dots};
\node       (q24) at (1.8,-2.1) {};
\node       (q13) at (1.8,2.1)  {};
\draw (0,0) node[rotate=90]{\footnotesize \dots\dots~~replicas~~\dots\dots};
\node[draw] (u1) at (2.1,0.7) {\footnotesize\emph{update detector}};
\node       (u11) at (3.1,0.7) {};
\node[draw] (u2) at (2.1,-0.7) {\footnotesize\emph{update detector}};
\node       (u22) at (3.1,-0.7) {};
\node[draw,rotate=90] (R) at (5.3,0) {\footnotesize\emph{~~~reconciler~~~}};
\node       (ap) at (5.3,2.2) {};
\node       (bp) at (5.3,-2.1) {};
\node         (cr) at (6.8,0) {};
\path[->]
      (q0) edge (q1) edge (q2)
      (q5) edge (q7) (q6) edge (q7)
      (q1) edge node[fill=white]{$\ss\alpha_0$} node[above=3pt]{\footnotesize updates} (q3)
      (q2) edge node[fill=white]{$\ss\beta_0$} node[below=3pt]{\footnotesize updates} (q4)
      (q3) edge node[fill=white]{$\ss\alpha_1$} node[above=3pt]{\footnotesize synchronization} (q5)
      (q4) edge node[fill=white]{$\ss\beta_1$} node[below=3pt]{\footnotesize synchronization} (q6);
\path[dotted,->]
      (q13) edge (u1) (q24) edge (u2);
\path[thick,->]
      (q2) edge (u2) (q4) edge (u2) 
      (q1) edge (u1) (q3) edge (u1)
      (u11) edge node[fill=white]{$\ss\alpha$} (R)
      (u22) edge node[fill=white]{$\ss\beta$} (R)
      (R) edge (ap) edge (bp); 
\end{tikzpicture}%
\caption{\fontsize{9}{10}\selectfont
Outline of the synchronization process.
Identical replicas of the original filesystem $\FS$ are updated (modified) yielding the
divergent replicas $\FS_1$ and $\FS_2$. The \emph{reconciler} uses
the update information $\alpha$ and $\beta$ extracted by the \emph{update
detector}s, and generates the synchronizing instructions $\alpha_1$ and
$\beta_1$. These create the identical merged state
$\FSa$ when applied to the replicas.
The update detectors determine the update information $\alpha$ and $\beta$
either by comparing the different states of the replicas (e.g. $\FS_1$ vs.
$\FS$), or by having access to the update instructions $\alpha_0$ and
$\beta_0$ that were applied to $\FS$.
}\label{fig:sync-process}
\end{figure}
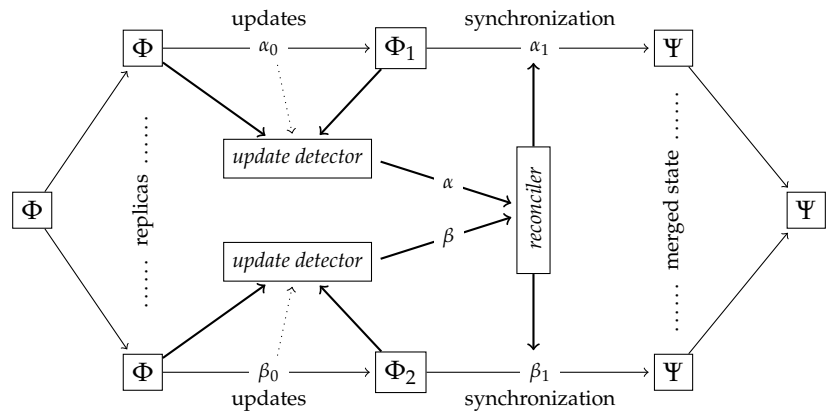

Both OT and CRDT have been successfully applied in a variety of
synchronization tasks \cite{NS16,TSR15}. Filesystem synchronization,
however, fits very poorly into these frameworks, mainly because it works
under a completely different \emph{modus operandi}: constant, low latency
communication in OT and CRDT versus a single data exchange in filesystem
synchronization; frequent loose synchronization with eventual convergence
requirement versus a single but complete synchronization; small number of
differences versus significant structural differences accumulated during an
extended time period; and so on. Consequently, for a theoretical
investigation of filesystem synchronization, we follow a traditional
framework instead, described in e.g. \cite{BP98} and depicted in Figure
\ref{fig:sync-process} adapted from \cite{CC22}.
Two identical replicas of the original filesystem $\FS$ are updated
independently, yielding the divergent replicas $\FS_1$ and $\FS_2$. In the
state-based case the \emph{update detector} receives the original ($\FS$)
and the current ($\FS_1$ or $\FS_2$) states, and generates the update
information $\alpha$ (or $\beta$) describing the differences between the
original filesystem and the replica. In the operation-based case the update
decoder has access to the performed operations $\alpha_0$ ($\beta_0$) only.
The \emph{reconciler} receives the information provided by the update
detectors and generates the synchronizing instructions $\alpha_1$ and
$\beta_1$, respectively. These instructions, applied to the replicas $\FS_1$
and $\FS_2$, create the identical \emph{merged state $\FSa$}.

In order to reach such a common synchronized state, existing theoretical and
practical filesystem synchronizers, such as
\cite{IceCube,MPV,PV,syncpal-thesis,TSR15,TTPDSH}, apply some
conflict-resolution strategy. They identify all (or some) of the conflicts,
then apply their strategy to the conflicts one at a time. These algorithms
are typically fixed and defined by some rationale or heuristics, or dictated
by the underlying technology. In contrast, we \emph{define what comprises a
synchronized state.} Intuitively, it is a maximal consistent merger of the
replicas, meaning that no further changes can be applied to the merged state
from those that were applied to the replicas during the update phase. Then
we proceed to prove that every such maximal state (and only these states)
can be reached by resolving the conflicts in some order.

In the meticulous survey \cite{syncpal-thesis}
of existing theoretical and practical filesystem
synchronizers it has been observed that ``resolving
conflicts is an open problem where [{\dots}] most academic works present
arbitrary resolution methods that lack a rationale for their decisions''
(page i), and that ``a file system can be affected by more than one conflict
at once, which has not been discussed in related academic works'' (page 65).
By presenting, for the first time, a complete analysis of the
synchronization process for two replicas of a theoretical filesystem model,
we fill these gaps.

\subsection{Our contribution}

This paper is a complete analysis of the synchronization process of two
replicas of a theoretical filesystem. Our filesystem model together
with the commands considered -- such as \emph{create}, \emph{rmdir}, or
\emph{edit} -- are discussed in Secion \ref{sec:definitions}. The changes
between the original replica and the locally updated versions are captured by
a special command set as defined in Section
\ref{sec:canonical}, which also defines algorithms generating this
update information.

The important question of which filesystems can, in general, be considered
to be a synchronized state of two divergent replicas is tackled in Section
\ref{sec:reconciler}. Our definition captures this notion in its full generality
without prescribing how the synchronized state can be reached. Providing
such a declaration-based definition of the synchronized state is one of our
main contributions.
Section \ref{sec:conflict} presents the generic
synchronization algorithm based on conflict resolution. By using different
strategies to resolve conflicts,
any of the
synchronized states allowed by our definition, and only those, can be the
result of the algorithm. 
Finally, Section \ref{sec:conclusion} summarizes
the results and lists open problems and directions for future research.

\section{Methodology}\label{sec:methods}

Our filesystem model, defined in Section \ref{sec:definitions}, is arguably
simple, but it retains all important structural properties of real
filesystems. It is this simplicity that allows us to exploit a rich and
intriguing algebraic structure \cite{Csi16,CC22} which eventually leads to
the claimed theoretical results. In Section \ref{sec:conclusion} we discuss
some possible extensions of the filesystem model.

Depending on the data communicated by the replicas, synchronizers are
categorized as either \emph{state-based} or \emph{operation-based}
\cite{syncpal-thesis,CRDT-overview}. In state-based synchronization replicas
send their current versions of the filesystem, or merely the differences
(called \emph{diff}s) between the current states and the last known
synchronized state \cite{AC08}. Operation-based synchronizers transmit the
complete log (or trace) of all operations performed by the user
\cite{IceCube}. The synchronization method of this paper is reminiscent of
operation-based one, but can also be considered, with similar overhead, to
be state-based. A set of virtual filesystem operations (commands) will be
defined in Section \ref{sec:definitions}. Each of these commands have a
clear and intuitive operational meaning, but are not necessarily available
to the end-user. The current state of the filesystem is described by a
special -- called \emph{canonical} -- sequence of virtual commands, which
transforms the original filesystem to the actual replica. The \emph{update
detector} can generate this canonical sequence from the operations performed
by the user on the replica (operation-based) as well as from the differences
between the original and final state (state-based). On Figure
\ref{fig:sync-process} these sequences correspond to the information in
$\alpha$ and $\beta$ passed to the reconciler.

Filesystem synchronization must resolve all conflicts between the replicas.
Conflict resolution, however, should be ``intuitively correct,'' i.e.
discarding all changes made by both replicas is not a viable alternative.
The majority of commercially or theoretically available synchronizers do not
present a rationale to explain their concrete conflict resolution approach
\cite{syncpal-thesis}. Two notable exceptions are \cite{TSR15} and
\cite{NS16} which describe high-level consistency philosophies. In
\cite{TSR15} the main principles are 1) \emph{no lost update:} preserve all
updates on all replicas because these updates are equally valid; and 2)
\emph{no side effects:} such as a merge where objects unexpectedly
disappear. While these principles intuitively make sense, it is easy to see
that neither could possibly be upheld for every conflict; even the authors
provide counterexamples. In \cite{NS16} the relevant consistency
requirements are worded as follows:
\begin{itemz}
\item[R1]\hypertarget{R1}{}
\emph{intention-confined effect:} operations applied to the replicas by the
synchronizer must be based on operations generated by the end-user; and 
\item[R2]\hypertarget{R2}{}
\emph{aggressive effect preservation:} the effect of compatible operations should
be preserved fully; and the effect of conflicting operations should be preserved 
as much as possible.
\end{itemz}
These requirements are in fact variations of the OT consistency model, see
for example the notion of intention preservation in \cite{SE98}. (We note
that the other two OT principles -- convergence and causality preservation
-- do not apply to filesystem synchronizers.)

In agreement with \hyperlink{R1}{R1} and \hyperlink{R2}{R2} we 
\emph{declare} what a synchronized state is,
rather than present an algorithm which generates it. Our declaration can be
outlined as follows. Suppose that the two replicas $\FS_1$ and $\FS_2$ are
represented by the canonical sequences $\alpha$ and $\beta$, respectively,
that is, $\FS_1=\alpha\FS$ and $\FS_2=\beta\FS$, where $\alpha\FS$ is the
result of applying the command sequence $\alpha$ to $\FS$. The
\emph{synchronized} or \emph{merged state $\FSa$} is determined by the
canonical sequence $\mu$ as $\FSa=\mu\FS$ such that
\begin{itemz}
\item[C0]\hypertarget{C0}{}
 $\mu$ is applicable to the original filesystem $\FS$,
\item[C1]\hypertarget{C1}{}
 every command in $\mu$ can be found either in $\alpha$, in $\beta$, 
or in both, and
\item[C2]\hypertarget{C2}{}
 $\mu$ is maximal, i.e. no canonical sequence adding more
commands to $\mu$ can satisfy both C0 and C1.
\end{itemz}
Condition \hyperlink{C0}{C0} is the obvious requirement that the synchronizer must not cause
errors. Condition \hyperlink{C1}{C1} ensures that the synchronization satisfies the
\emph{intention-confined effect} (no surprise changes in the merged
filesystem) requirement \hyperlink{R1}{R1} as $\mu$ should consist only of commands which
were supplied by the replicas. The other consistency requirement
\hyperlink{R2}{R2} is
guaranteed by \hyperlink{C2}{C2} as $\mu$ is maximal, therefore it preserves as much of the
intention of the users as possible. Note that this definition of a
synchronized state is never vacuous. There are only finitely many sequences
which satisfy \hyperlink{C0}{C0} and \hyperlink{C1}{C1} 
as every command in $\mu$ comes from either
$\alpha$ or $\beta$ and no repetitions are allowed,
and there are at least two, namely $\alpha$ and
$\beta$. Because of this, the empty sequence (undoing all changes) will not
satisfy \hyperlink{C2}{C2} (assuming one of $\alpha$ and $\beta$ is not empty), in line with
our requirements.

The declaration-based synchronization is intuitively clear, easy to
understand, and does not use any predetermined conflict-resolving policy. An
operational characterization is proved in Theorem \ref{thm:merge}. The
essence of this theorem is that any merged filesystem $\FSa$ can be
generated from the replica, say $\FS_1$, by
\begin{itemz}
\item[1.] rolling back some of the commands in $\alpha$, followed by
\hfill\hypertarget{property}{\raisebox{-5pt}[0pt][0pt]{(1)}}
\item[2.] applying some commands from the other sequence $\beta$,
\end{itemz}
and \emph{vice versa} for the other replica. The commands rolled back
represent a minimal set whose removal resolves all conflicts. These commands
also give users a clear understanding of the changes the synchronizer wants
to perform on their filesystem. They are also helpful when some of the
rolled-back commands should be introduced again after the merging (doing a
redo of the undo).

Turning to traditional conflict-based synchronization, Section
\ref{sec:conflict} proves that any dec\-la\-ra\-tion-based merged state, and
only those states, can be the result of a conflict-based synchronization
algorithm. For each pair of canonical sequences describing the replicas to
be synchronized we define what the conflicting command pairs are. Their
resolution uses the winner/loser paradigm: the winner command is accepted
while the loser command is discarded. It turns out that resolving a conflict
may automatically resolve some of the subsequent conflicts, but no new
conflicts are created. Using this result the outcomes of different conflict
resolving policies can be investigated easily. In particular, the
\emph{iterative approach} \cite{syncpal-thesis} always works, which applies
all non-conflicting commands, resolves, in any way, the first conflict
(which might automatically resolve other existing conflicts), and iterates.

\section{Definitions}\label{sec:definitions}

This section defines the basic notation which will be used throughout the
rest of the paper. The exposition follows \cite{Csi16} and \cite{CC22} with
some substantial modifications. Some results from these papers are included
without proof.

\subsection{Namespace and filesystems}\label{subsec:filesystem}

Our filesystem model is arguably simplistic, nevertheless it captures all
important aspects of real-word implementations. In spirit, it is a mixture
of \emph{identity-} and \emph{path-based} models
\cite{syncpal-thesis,TSR15}. Objects are stored in nodes, which are uniquely
identified by fixed and predetermined \emph{path}s. The set of available
nodes is fixed in advance, and no path operations are considered.
Filesystems are required to have the \emph{tree-property} at all times: in a
given filesystem along any branch starting from any of the root nodes, there
must be zero or more \emph{directories}, zero or one \emph{file}, followed
by \emph{empty} nodes. Our model does not support \emph{links} (but see
Section \ref{subsec:links}), thus the namespace -- the set of available
nodes or paths -- forms a collection of rooted trees. Filesystems are
defined over and populate this fixed namespace.

Formally, the \emph{namespace} is a set $\N$ endowed with the partial
function $\parent:\N\to\N$ returning the parent of every non-root node (it
is not defined on roots). If $n=\parent m$ then $n$ is the parent of $m$,
and $m$ is a child of $n$. For two nodes $n,m\in\N$ we say that \emph{$n$ is
above $m$}, or $n$ is an ancestor of $m$, and write $n\prec m$ if
$n=\parent^i(m)$ for some $i\ge 1$. As usual, $n\preceq m$ means $n\prec m$
or $n=m$. As the parent function induces a tree-like structure, $\preceq$ is
a partial order. Two nodes $n,m\in\N$ are \emph{comparable} if either
$n\preceq m$ or $m\preceq n$, and they are \emph{uncomparable} or
\emph{independent} otherwise.

A \emph{filesystem $\FS$} populates the nodes of the namespace with values.
The value stored at node $n\in\N$ is denoted by $\FS(n)$. This value can be
$\E$ indicating that the node is \emph{empty} (no content, not to be
confused with the empty file); can be $\D$ indicating that the node is a
\emph{directory}; otherwise it is a \emph{file} storing the complete content
(which could happen to be ``no content''). We use $\E$, $\D$ and $\F$ to
denote the \emph{type} of the content corresponding to these possibilities.
While there is only one value of type $\E$ and one value of type $\D$ (see
Section \ref{subsec:attributes} for a discussion on lifting this
limitation), there are many different file values of type $\F$ representing
different file contents.

\subsection{Internal filesystem commands}\label{subsec:commands}

The synchronizer operates using a specially designed and highly symmetrical
set of \emph{internal filesystem commands}. They each modify the filesystem
at a single node only, and contain additional information usually not
thought of as part of a command which allows them to be inverted, and makes
them amenable to algebraic manipulation. Inventing such a command set was
one of the main contributions of \cite{Csi16}.

The commands basically implement creating and deleting files and
directories. Modifying a part of a file only is not available as a command;
whenever a file is modified, the new content must be supplied in its
entirety. Still, commands issued by a real-life user or system can be easily
transformed into the internal commands; for example, the \emph{move} or
\emph{rename} operation can be modeled as a sequence of a \emph{delete} and
a \emph{create}.

The set of the internal commands is denoted by $\Omega$, and each command
$\sigma\in\Omega$ has three components, written as $\sigma=\(n,x,y)$, as
follows:
\begin{itemz}
\item $n\in\N$ is the node on which the command acts,
\item $x$ is the content at node $n$ before the command is executed
(precondition), and
\item $y$ is the content at node $n$ after the command was executed.
\end{itemz}
Thus \emph{rmdir}$(n)$ corresponds to the internal command $\(n,\D,\E)$,
which replaces the directory value at $n$ by the empty value. The command
$\(n,\E,\D)$ creates a directory at $n$, but only if the node $n$ has no
content, i.e., there is no directory or file at $n$ (a usual requirement for
\emph{mkdir}$(n)$). For files $\f_1$ and $\f_2\in\F$ the command
$\(n,\f_1,\f_2)$ replaces $\f_1$ stored at node $n$ by the new content
$\f_2$. This latter command can be considered to be an equivalent of
\emph{edit}$(n,\f_1,\f_2)$.

Applying $\sigma\in\Omega$ to a filesystem $\FS$ is written as the left
action $\sigma\FS$. The command $\sigma=\(n,x,y)$ is \emph{applicable} to
$\FS$ if
\begin{itemz}
\item $\FS$ contains $x$ at the node $n$, that is, $\FS(n)=x$, and
\item after changing the content at $n$ to $y$ the filesystem still has the
tree property.
\end{itemz}
If $\sigma$ is not applicable to $\FS$ then we say that $\sigma$
\emph{breaks} the filesystem, and write $\sigma\FS=\bot$. If $\sigma$ does
not break $\FS$, then $\sigma\FS$ is the filesystem where every node has the
same value as in $\FS$ except for the node $n$ where the new content is $y$.
Command sequences are applied from left to right, thus $(\sigma\alpha)\FS =
\alpha(\sigma\FS)$. The composition of sequences $\alpha$ and $\beta$ is
written as $\alpha\beta$; occasionally we write it as $\alpha\circ\beta$ to
emphasize that $\beta$ is to be executed after $\alpha$. A sequence breaks
$\FS$ if one of its commands is not applicable. More formally,
$\sigma\alpha$ breaks $\FS$ if either $\sigma$ breaks $\FS$, or $\alpha$
breaks $\sigma\FS$.

Two additional commands will be used, which are denoted by $\eps$ and
$\lambda$, respectively. In practice they do not occur, and cannot be
elements of command sequences, but are useful in algebraic derivations. The
command $\eps$ breaks every filesystem, while $\lambda$ acts as identity:
$\eps\FS=\bot$ and $\lambda\FS=\FS$ for every filesystem $\FS$.

The command sequences $\alpha$ and $\beta$ are \emph{semantically
equivalent}, written as $\alpha\equiv\beta$ if they have the same effect on
all filesystems: $\alpha\FS=\beta\FS$ for all $\FS$. We write $\alpha\semle
\beta$ to denote that \emph{$\beta$ semantically extends $\alpha$}, that is,
$\alpha\FS=\beta\FS$ for all filesystems that $\alpha$ does not break.
Clearly, $\alpha\equiv\beta$ if and only if both $\alpha\semle\beta$ and
$\beta\semle\alpha$. The sequence $\alpha$ is \emph{non-breaking} if there
is 
at least one filesystem $\FS$ which
$\alpha$ does not break. This is the same as requesting
$\alpha\not\equiv\eps$.

The \emph{inverse} of $\sigma=\(n,x,y)$ is $\sigma^{-1}=\(n,y,x)$. For a
command sequence $\alpha$ its inverse is defined in the usual way:
$\alpha^{-1}$ consists of the inverse of the commands in $\alpha$ written in
reverse order. This inverse has the expected property: if $\alpha$ does not
break $\FS$, then $(\alpha\alpha^{-1})\FS = \FS$. In particular,
$\alpha\alpha^{-1} \semle\lambda$.

\subsection{Command types and execution order}

A node value has type $\E$, $\D$, or $\F$ depending on whether it is the
empty value, a directory, or a file. For a command $\sigma=\(n,x,y)$ the
type of $x$ is its \emph{input type} and the type of $y$ is its \emph{output
type}. Depending on their input and output types commands can be partitioned
into nine disjoint classes. To make their descriptions clearer, we use
\emph{patterns}. The command $\(n,x,y)$ matches the pattern $(n,\mathsf
P_x,\mathsf P_y)$ if the type of $x$ is listed in $\mathsf P_x$ and the
type of $y$ is listed in $\mathsf P_y$. In a pattern the symbol $\bullet$
matches any value. As an example, every command matches the pattern
$\(\bullet,\E\F\D,\D\F\E)$.

\emph{Structural commands} change the type of the stored data, while
\emph{transient commands} retain it. In other words, commands matching
$\(\bullet,\E,\F\D)$, $\(\bullet,\F,\E\D)$ or $\(\bullet,\D,\E\F)$ are
structural commands, while those matching $\(\bullet,\E,\E)$,
$\(\bullet,\F,\F)$ or $\(\bullet,\D,\D)$ are transient ones. Commands with
identical input and output values are \emph{null commands}. Every null
command is transient, and transient commands which are not null match the
pattern $\(\bullet,\F,\F)$. If $\sigma$ is a null command, then
$\sigma\semle\lambda$ meaning that $\sigma$ does not change the filesystem
(but can break it), and if $\sigma\semle\lambda$ then $\sigma$ is a null
command.

Structural commands are further split into \emph{constructors} and
\emph{destructors}. Constructor commands upgrade the type from $\E$ to $\F$
to $\D$, while destructors downgrade it. That is, constructors are commands
matching $\(\bullet,\E,\F\D)$ or $\(\bullet,\F,\D)$, while the destructors
match $\(\bullet,\D,\F\E)$ or $\(\bullet,\F,\E)$.

Some command pairs on parent--child nodes can only be executed in a certain
order. This notion is captured by the binary relation $\sigma\orel\tau$ with
the meaning that $\sigma$ must precede $\tau$ in the \emph{execution order}.

\begin{Definition}[Execution order]\label{def:ex-order}
For a command pair $\sigma,\tau\in\Omega$ the binary relation
$\sigma\orel\tau$ holds if the pair matches either $\(n,\D\F,\E) \orel
\(\parent n,\D,\F\E)$ or $\(\parent n,\E\F,\D) \orel \(n,\E,\F\D)$.
An \emph{$\orel$-chain} is a sequence of commands such that $\sigma_1 \orel
\sigma_2 \orel \cdots \orel \sigma_k$. An $\orel$-chain \emph{connects} its
first and last element.
\end{Definition}

\begin{figure}[H]
\begin{adjustwidth}{-\extralength}{0cm}\centering\hsize=\linewidth
\begin{tikzpicture}[text depth=-0.2ex,text height=1.2ex]
\begin{scope}[xshift=-9cm,yshift=-2.5cm]
\begin{scope}[yshift=3.5cm]
\Dfive\D\D\D\D{\D\F}
\draw (-2.5,0) node[right]{\footnotesize input type};
\end{scope}
\begin{scope}[yshift=2.5cm]
\Dfive{\F\E}\E\E\E\E
\RAB 1 \RAB2 \RAB 3 \RAB 4
\draw (-2.5,0) node[right]{\footnotesize output type};
\DBOX 0 \DBOX 1 \DBOX 2 \DBOX 3 \DBOX 4
\foreach\x in{1,...,5}{
  \draw ({(5-\x)*\MM},1.6) node {$\sigma_\x$};
}
\draw(-2.5,1.6) node[right]{\footnotesize command};
\end{scope}
\end{scope}
\begin{scope}[yshift=1.0cm]
\Dfive{\F\E}\E\E\E\E
\draw (-2.5,0) node[right] {\footnotesize input type};
\end{scope}
\Dfive\D\D\D\D{\D\F}
\RAF 0 \RAF 1 \RAF 2 \RAF 3
\draw (-2.5,0) node[right] {\footnotesize output type};
\DBOX 0 \DBOX 1 \DBOX 2 \DBOX 3 \DBOX 4
\foreach\x in{1,...,5}{
  \draw ({(\x-1)*\MM},1.6) node {$\sigma_\x$};
}
\draw(-2.5,1.6) node[right]{\footnotesize command};
\end{tikzpicture}%
\caption{\fontsize{9}{10.2}\selectfont
Structure of an up (left) and down (right) $\orel$-chain.
Black arrows point to the parent node and red arrows show the execution order 
$\orel$.
}\label{fig:orel-chain}
\end{adjustwidth}\end{figure}
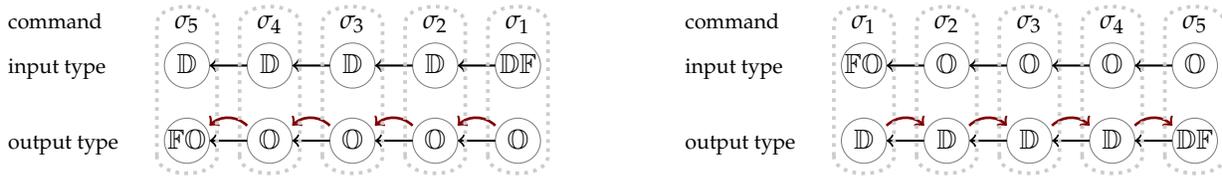

\noindent
The first case in Definition \ref{def:ex-order} of the $\orel$ relation
corresponds to the requirement that a directory cannot be deleted when its
descendants are not empty. The second case requires that before creating a
file or directory, the directory holding it should exist. Observe that
$\sigma \orel\tau$ implies that both $\sigma$ and $\tau$ are structural
commands on parent--child nodes, and either both are constructors or both
are destructors. It means that elements of an $\orel$-chain match either the
pattern
\begin{disp}
   \(n_1,\D\F,\E) \orel \(n_2,\D,\E) \orel
\cdots \orel \(n_{k-1},\D,\E) \orel \(n_k,\D,\F\E)
\end{disp}
where $n_{i+1}$ is the parent of $n_i$ (we move up along a branch), or they
match
\begin{disp}
   \(n_1,\F\E,\D) \orel \(n_2,\E,\D) \orel \cdots \orel \(n_{k-1},\E,\D)
     \orel \(n_k,\E,\D\F),
\end{disp}
where $n_i$ is the parent of $n_{i+1}$ (we move down). Figure
\ref{fig:orel-chain} illustrates the structure of $\orel$-chains. The
circles represent the nodes on which the commands act: the top, input type
row denotes the types of their values before executing the commands, while
the bottom, output type row shows the types of their values afterwards. The
black arrows point towards the parent nodes, and the read arrows indicate
the execution order.

\section{Canonical sequences and sets}\label{sec:canonical}

Our synchronizer, including the update detector, relies heavily on the
algebraic structure of internal command sequences. The theory of this
structure has been developed in \cite{Csi16} and \cite{CC22}. We provide
high-level overviews of the proofs whenever appropriate to make this paper
self-contained and also available to non-experts. For full proofs the reader
is referred to the original papers.

To understand the semantical properties of command sequences, the first step
is to investigate command pairs and determine when they commute, and when
they require a special execution order. Proposition \ref{prop:1} covers the
case when the commands act on the same node, and Proposition \ref{prop:2}
considers command pairs on different nodes.

\begin{Proposition}[{\cite[Proposition 4]{CC22}}]\label{prop:1}
Suppose $\sigma,\tau\in\Omega$ are on the same node $n$. Then one of the
following two possibilities hold:
\begin{itemz}[0pt]
\item[(a)] $\sigma\tau\semle \omega$ for some $\omega\in\Omega$ also on the
same node $n$,
\item[(b)] $\sigma\tau\equiv\eps$, that is, the pair breaks every filesystem.
\end{itemz}
\end{Proposition}

\begin{proof}
Suppose $\sigma=\(n,x_1,y_1)$ and $\tau=\(n,x_2,y_2)$. If $y_1=x_2$ then (a) 
holds with $\omega=\(n,x_1,y_2)$. If $y_1\neq x_2$ then (b) holds: if
$\sigma$ does not break the filesystem, then it leaves $y_1$ at node $n$.
Now $\tau$ requires $x_2$ there, but finds a different value, thus breaks the
filesystem.
\end{proof}

For two commands we write $\sigma \indep \tau$ to mean that the nodes they
act on are uncomparable, i.e., the nodes are different and neither is an
ancestor of the other.

Recall that $\sigma$ is a null command if it has the same input and output
value, that is, it is of the form $\sigma=\(n,x,x)$.

\begin{Proposition}[{\cite[Proposition 5]{CC22}}]\label{prop:2}
Suppose $\sigma$ and $\tau$ are non-null commands on different nodes.
Then
\begin{itemz}[0pt]
\item[(a)] if $\sigma\indep \tau$ then $\sigma\tau \equiv \tau\sigma$, and
\item[(b)] if $\sigma\notindep\tau$ then $\sigma\tau \not\equiv \eps$ if and only
if $\sigma \orel \tau$.
\end{itemz}
\end{Proposition}

\begin{proof}
By checking all possibilities. 
\end{proof}

\subsection{Canonical sequences}

Informally, canonical sequences are the ``clean'' versions of non-breaking
command sequences. The initial definition is a mixture of syntactical and
semantical properties. Proposition \ref{prop:3} defines an algorithm which
converts any sequence into a canonical one purification, and Theorem
\ref{thm:canonical} gives a complete syntactical characterization. An
important consequence of this characterization is that the semantics of a
canonical sequence is determined uniquely by the unordered set of commands
it contains. It means that the order of the commands -- up to semantical
equivalence -- can be recovered from their set only.


\begin{Definition}[Canonical sequences]\label{def:canonical}
A command sequence is \emph{canonical} if
\begin{itemz}[0pt]
\item[(a)] it does not contain null commands, that is, a command of the 
form $\(n,x,x)$,
\item[(b)] no two commands in the sequence are on the same node, and
\item[(c)] it is non-breaking.
\end{itemz}
\end{Definition}

\noindent
Conditions (a) and (b) are clearly syntactical, and can be checked by
looking at the commands in the sequence. In Theorem \ref{thm:canonical}
below we give a purely syntactical characterization of canonical sequences.
Before stating the theorem we show that canonical sequences are the
``clean'' versions of arbitrary command sequences.

\begin{Proposition}[{\cite[Theorem 27]{Csi16}}]\label{prop:3}
For every non-breaking command sequence $\alpha$ there is a
canonical sequence $\alpha^*$ such that $\alpha\semle\alpha^*$, that is, if
$\alpha$ does not break $\FS$, then both $\alpha$ and $\alpha^*$ 
have
the same effect on $\FS$.
\end{Proposition}

\begin{proof}
Null commands can be deleted from $\alpha$ as they do not change the
filesystem (but may break it). Suppose $\alpha$ has two commands on the same
node. Choose two which are closest to each other; let these be $\sigma_1$
and $\sigma_2$, both on node $n$. The output value of $\sigma_1$ must be the
same as the input value of $\sigma_2$ (as $\alpha$ is non-breaking and the
value at node $n$ does not change between $\sigma_1$ and $\sigma_2$).
If $\sigma_1$ and $\sigma_2$ are not next to each other, then using
Proposition \ref{prop:2}, either $\sigma_1$ can be moved forward by swapping
it with the immediately following command (which, by assumption, is on a
different node), or $\sigma_2$ can be moved backward by swapping it with the
immediately preceding command. If none of those swaps can be done, then
$\alpha$ would break every filesystem. Eventually $\sigma_1$ and
$\sigma_2$ can be moved next to each other. In this case, however, Proposition
\ref{prop:1} implies that $\sigma_1\sigma_2$ can be replaced by a simgle
command.
\end{proof}

\noindent
This proof is constructive and specifies a quadratic algorithm which
transforms any non-breaking sequence into a canonical one. To state the
purely syntactical characterization of canonical sequences we need further
definitions.

\begin{Definition}
Let $\alpha$ be a command sequence.
\begin{itemz}[0pt]
\item[(a)] $\alpha$ \emph{honors $\orel$}, if for
commands $\sigma,\tau\in\alpha$, $\sigma$ precedes $\tau$ whenever $\sigma 
\orel \tau$.
\item[(b)] $\alpha$ is \emph{$\orel$-connected}, if for any two commands
$\sigma,\tau\in\alpha$, either $\sigma \indep \tau$, or $\sigma$ and $\tau$
are connected by an $\orel$-chain (see Definition \ref{def:ex-order}) whose
elements are in $\alpha$.
\end{itemz}
\end{Definition}

\begin{Theorem}[Syntactical characterization of canonical sequences]\label{thm:canonical}
A command sequence is canonical if and only if
\begin{itemz}[0pt]
\item[(a)] it does not contain null commands,
\item[(b)] no two commands in the sequence are on the same node,
\item[(c1)] it honors $\orel$, and
\item[(c2)] it is $\orel$-connected.
\end{itemz}
\end{Theorem}

\begin{proof}
Let $\alpha$ be a canonical sequence according to Definition
\ref{def:canonical}. To see that if either (c1) or (c2) fails then $\alpha$
breaks every filesystem, use induction on the length of $\alpha$. This shows
that if $\alpha$ contains commands on nodes $n$ and $m$ which lie on the
same branch, then $\alpha$ must contain commands on every node between $n$
and $m$ in the right order.

For the reverse direction assume $\alpha$ satisfies conditions (a), (b),
(c1) and (c2). We create a filesystem which $\alpha$ does not break. The
values at nodes mentioned in the sequence are set to the input values of the
commands. Values at nodes that are ancestors of the ones mentioned in the
sequence are set to directories. Other nodes of the filesystem remain empty.
This is a valid filesystem, and by property (c2) every command in $\alpha$
has the correct input value. Using property (c1) one can prove that $\alpha$
works on this filesystem.
\end{proof}

\subsection{Canonical sets}\label{subsec:csets}

An important consequence of Theorem \ref{thm:canonical} is that the
semantics of a canonical sequence is uniquely determined by the commands it
contains. This is expressed formally in the following Proposition.

\begin{Proposition}[{\cite[Theorem 14]{CC22}}]\label{prop:canonical-set}
If the canonical sequences $\alpha$ and $\beta$ contain the same
commands, then they are semantically equivalent, that is, $\alpha\FS=\beta
\FS$ for every filesystem.
\end{Proposition}
\begin{proof}
In fact, $\alpha$ and $\beta$ can be transformed into each other using the
commutativity rules given in Proposition \ref{prop:2}(a), see \cite[Lemma
10]{CC22}.
\end{proof}

\begin{Definition}[Canonical set]\label{def:canonical-set}
The set $A$ of internal commands is \emph{canonical} if
\begin{itemz}[0pt]
\item[(a)] it does not contains null commands,
\item[(b)] no two commands in $A$ are on the same node, and
\item[(c2)] the set $A$ is $\orel$-connected, meaning that if two of its
commands are on comparable nodes, then they are connected by an
$\orel$-chain whose elements are in $A$.
\end{itemz}
\end{Definition}

Commands in a canonical sequence form a canonical set by Theorem
\ref{thm:canonical}. In the other direction, a canonical set can be ordered
in many ways to become a canonical sequence -- the only requirement is (c1)
which requires that the order honors the relation $\orel$. This can be
achieved by using, e.g., topological sort. Any two such orderings give
semantically equivalent sequences, as was proved in Proposition
\ref{prop:canonical-set}. Consequently we can, and will, use canonical sets
in places where command sequences are required with the implicit
understanding that the commands in it are ordered in an appropriate fashion.
For example, we write $A\FS$ to mean $\alpha\FS$ where $\alpha$ contains the
elements of $A$ in an order which honors $\orel$.

Canonical sets play an essential role in reconciliation (Section
\ref{sec:reconciler}) and conflict resolution (Section \ref{sec:conflict}).
One of their crucial properties is, as we have seen, that their commands can
be executed in different orders while preserving their semantics -- a
variant of the \emph{commutativity principle} of CRDT \cite{PMS09}.
Proposition \ref{prop:5} discusses the special case when a subset of the
commands should be moved to the beginning of the execution line. Then we
define and investigate \emph{clusters} in a canonical set, and observe that
the clusters can be freely rearranged, and even intertwined.

\begin{Definition}\label{def:subc}
For a canonical set $A$ we write $B \subc A$ to indicate that $B$ is not
only a subset of $A$ but can also be moved to the beginning while keeping the
semantics, namely $A \equiv B \circ (A\setm B)$.
\end{Definition}

\begin{Proposition}\label{prop:5}
Let $A$ be a canonical set and $B\subseteq A$. Then $B\subc A$ if and only
if $\sigma\in A$, $\tau \in B$ and $\sigma \orel \tau$ imply $\sigma\in B$. If
$B\subc A$ then both $B$ and $A\setm B$ are canonical.
\end{Proposition}

\begin{proof}
An ordering of $A$ does not break every filesystem if and only if it
honors $\orel$, see \cite[Lemma 10]{CC22}. Consequently, if 
$A\equiv B \circ (A\setm B)$ then this split must honor $\orel$, which is 
exactly the stated condition. If this condition holds, then $A$ has an
ordering which honors $\orel$ and in which $B$ precedes $A\setm B$.

For the second part it suffices to check that both $B$ and $A\setm B$ are
$\orel$-connected. But it is immediate from the condition and from the fact
that $A$ is $\orel$-connected.
\end{proof}

The $\orel$ relation connects commands in a canonical set $A$. The
\emph{clusters} of $A$ are the components of this connection graph. All
commands in a cluster are constructors, or all commands are destructors, or
the cluster has a single element matching $\(\bullet,\F,\F)$. Accordingly,
we call the cluster \emph{constructor}, \emph{destructor}, or \emph{editor},
respectively. Nodes of the commands within a cluster form a connected
subtree of the namespace $\N$, and the topmost elements of the subtrees are
pairwise incomparable. In a constructor cluster the ``being the parent'' and
the $\orel$ relations coincide, while in a destructor cluster they are the
reverse of each other, see Figure \ref{fig:orel-chain}. As feasible orders
of the commands in $A$ must only honor the relation $\orel$, clusters can
freely move around. In a constructor cluster a command can be executed only
\emph{after} all commands on its ancestors have been executed. In a
destructor cluster it is the other way around: a command can be executed
only after all commands on its descendants have been executed. Editor
commands form single-element clusters, thus they can be executed at any
point.

\subsection{The update detector}

Algorithms creating the canonical sets required by the reconciler (see
Figure \ref{fig:sync-process}) are described in Propositions \ref{prop:6}
and \ref{prop:7}. The update detector 
is either \emph{state-based}, in which case it has access to the original and 
modified filesystems, 
or it is \emph{command-based}, in which case it has access to the exact
sequence of commands used to modify the filesystem. Both update
algorithms are highly effective; essentially
their runtime is
linear in the input size.

\begin{Proposition}[{State-based update detector, \cite[Theorem 19]{CC22}}]\label{prop:6}
Let $\FS$ be the original filesystem and let $\FS_1$ be the replica. The
command set
\begin{disp}
  A^*= \{ \(n,\FS(n),{\FS_1(n)}) : n\in\N \mbox{~ and ~}
      \FS(n)\neq\FS_1(n)\,\}
\end{disp}
is canonical, and $A^*\FS = \FS_1$.
\end{Proposition}

\begin{proof}
To show that $A^*$ is canonical we need to check that it is
$\orel$-connected. By \cite[Lemma 23]{Csi16} if $n\prec m$ and $\FS$ and
$\FS_1$ differ at $n$ and at $m$, then they must differ at every node
between $n$ and $m$. Moreover, the differences follow one of the the patterns
on Figure \ref{fig:orel-chain}, thus these commands form the required
$\orel$-chain. A consequence of this is that $A^*$ does not break $\FS$, and
since after executing the commands every node will have the output value
from $\FS_1$, we know $A^*\FS=\FS_1$.
\end{proof}

The set $A^*$ can be generated by performing simultaneous depth-first searches on
the visible (non-empty) parts of the filesystems $\FS$ and $\FS_1$. Such an
algorithm runs in constant space, and the running time is linear in the total
size of the visible parts of the filesystems.

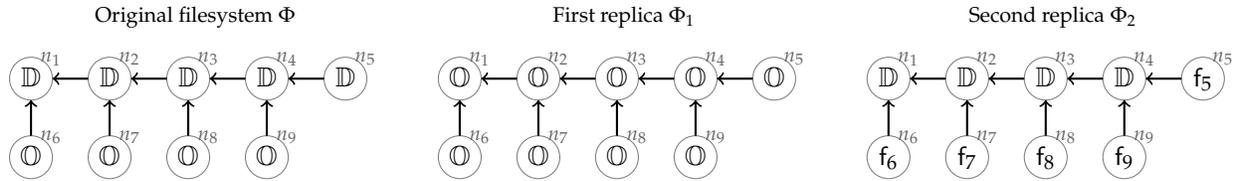
\begin{figure}[H]
\begin{adjustwidth}{-\extralength}{0cm}\centering\hsize=\linewidth
\begin{tikzpicture}[scale=0.95,text depth=-0.2ex,text height=1.2ex]
\draw(2.3,0.9) node {\footnotesize Original filesystem $\FS$};
\Dnine\D\D\D\D\D\E\E\E\E
\begin{scope}[xshift=6cm]
\draw(2.3,0.9) node {\footnotesize First replica $\FS_1$};
\Dnine\E\E\E\E\E\E\E\E\E
\end{scope}
\begin{scope}[xshift=12cm]
\draw(2.3,0.9) node {\footnotesize Second replica $\FS_2$};
\Dnine\D\D\D\D{\f_5}{\f_6}{\f_7}{\f_8}{\f_9}
\end{scope}
\end{tikzpicture}%
\caption{\fontsize{9}{10.2}\selectfont
Illustrating the update detector. The nodes are, from left to
right, $n_1$ to $n_5$ in
the top row, and $n_6$ to $n_9$ in the bottom row. The first replica deletes
all directories, the second replica stores file content $\f_i$ at node $n_i$.
}\label{fig:update}
\end{adjustwidth}\end{figure}

Figure \ref{fig:update} provides two examples for the result of the
state-based update detector. The namespace $\N$ consists of the nodes $n_1$
to $n_9$. The original filesystem $\FS$ contains directories at nodes $n_1$
to $n_5$ and the empty value at $n_6$ to $n_9$. The first replica deletes
all directories, setting the empty value everywhere. The second replica
stores file contents in $n_5$ and in the empty nodes. The canonical command
set transforming $\FS$ into $\FS_1$ consists of the five commands
$A=\{\sigma_1,\sigma_2,\sigma_3,\sigma_4,\sigma_5\}$ where $\sigma_i =
\(n_i,\D,\E)$ corresponding to the five differences between $\FS$ and
$\FS_1$. The second command set also has five commands
$B=\{\tau_5,\tau_6,\tau_7,\tau_8,\tau_9\}$ with $\tau_5=\(n_5,\D,\f_5)$ and
$\tau_i=\(n_i,\E,\f_i)$ for $i>5$. The set $A$ has a single execution order
honoring $\orel$, namely $\sigma_5$, $\sigma_4$, $\sigma_3$, $\sigma_2$,
$\sigma_1$. Any ordering of the commands in $B$ honors $\orel$, as no two
commands in $B$ are $\orel$-related.

\begin{Proposition}[Command-based update detector]\label{prop:7}
Let $\alpha$ be a command sequence that transforms the original filesystem
$\FS$ into the replica $\FS_1$. The following procedure creates a canonical
set $A^*$ for which $A^*\FS=\alpha\FS=\FS_1$. Let $A^*$ be empty initially,
and iterate over the commands in $\alpha$ from left to right. Let the next
command be $\sigma$. If there is no command in $A^*$ on the same node as
$\sigma$, add $\sigma$ to $A^*$. If $\sigma^*\in A^*$ and $\sigma$ are on
the same node, then using Proposition \ref{prop:1}(a) replace $\sigma^*$
with the single command corresponding to the composition $\sigma^*\sigma$.
Finally, when the iteration ends, delete the null-commands from $A^*$.
\end{Proposition}

\begin{proof}
It is clear that on each node of the filesystem the sequence
$\alpha$ and the set $A^*$ 
have
the same effect. Consequently $A^*$ is
equivalent to the command set created in Proposition \ref{prop:6}, therefore
it is canonical.
\end{proof}

If a hash value of the nodes is also stored, the algorithm can be implented so
that its time and space complexity is linear in the length of sequence
$\alpha$.

\section{The reconciler -- synchronizing two replicas}\label{sec:reconciler}

The update detector -- either state-based or operation-based -- passes the
canonical sets $A$ and $B$ to the reconciler, see Figure
\ref{fig:sync-process}. These sets describe, in a straightforward and
intuitive way, how the replicas $\FS_1$ and $\FS_2$ relate to the original
filesystem $\FS$.
As discussed in Section \ref{sec:methods}, the merged filesystem is also
specified by a canonical set $M$ (that of the sequence $\mu$) to be applied
to the original $\FS$. The next definition, which is the formalization of
the informal discussion 
in Section \ref{sec:methods}
about synchronization, \emph{declares} when 
the command set
$M$
represents an intuitively correct synchronization.

\begin{Definition}[Merger command set]\label{def:merger}
The \emph{merger} of the canonical sets $A$ and $B$ is a maximal canonical
set $M\subseteq A\cup B$, meaning that no proper superset of $M$ within
$A\cup B$ is canonical.
\end{Definition}

\noindent
Note that the merger $M$ is not necessarily unique, and typically many
different mergers exist (see the example below). This definition covers
every permissible synchronization outcome which satisfies both the
\emph{intention-confined effect} 
\hyperlink{R1}{R1}
($M$ is a subset of $A\cup B$) and the
\emph{aggressive effect preservation}
\hyperlink{R2}{R2}
($M$ is maximal) requirements
as discussed in Section \ref{sec:methods}.
When
performing the actual file synchronization, one of the mergers should be
chosen, either by the system or the user. Section \ref{sec:conflict}
discusses how this choice can be made by successive conflict resolutions.

In this section we prove that every synchronized state defined by a merger
$M$ has a clear operational characterization: both replicas can be
transformed into the merged state by first rolling back some of the commands
executed earlier on this replica, and then applying some commands executed
on the other replica.
Using the terminology of \cite{CC22}, the canonical sets $A$ and $B$ are
called \emph{refluent} if there is filesystem on which 
both of them work
(neither of them breaks). This condition is clearly met when $A$ and $B$
describe two replicas, as both are applied to their common original
filesystem.

\begin{Theorem}\label{thm:merge}
Let $A$ and $B$ be refluent canonical sets and $M \subseteq A \cup
B$ be a merger. Then
\begin{itemz}[2pt]
\item[(a)] $M\semge A\circ A_1^{-1}\circ B'$, where $A_1=A\setm M\subseteq A$ and
$B'=M\setm A\subseteq B$,
\item[(b)] $M\semge B\circ B_1^{-1}\circ A'$ where $B_1=B\setm M\subseteq B$ and
$A'=M\setm B\subseteq A$,
\item[(c)] if both $A$ and $B$ are applicable to a filesystem, then so
are $M$, $A \circ A_1^{-1} \circ B'$ and $B \circ B_1^{-1} \circ A'$.
\end{itemz}
\end{Theorem}

In case of synchronizing the replicas $\FS_1=A \FS$ and $\FS_2=B\FS$, let
$\FSa=M\FS$ be (one of) the merged 
filesystem(s). 
According to this theorem,
we have
\begin{disp}\begin{array}{l}
 \FSa=M\FS = (A\circ A_1^{-1}\circ B')\FS =
B'(A_1^{-1}\FS_1), \mbox{~~ and} \\[4pt]
 \FSa=M\FS = (B\circ B_1^{-1}\circ A')\FS =
A'(B_1^{-1}\FS_2).
\end{array}
\end{disp}
The first line says that the replica $\FS_1$ can be transformed into the
synchronized filesystem $\FSa$ first by applying $A_1^{-1}$, that is,
rolling back those commands in $A$ which are not in $M$, and then applying
some further commands $B'$ from $B$. The similar statement for the other
replica $\FS_2$ follows from the second line. This justifies the informal
statement (\hyperlink{property}{1})
in Section \ref{sec:methods}.


\begin{figure}[H]
\begin{adjustwidth}{-\extralength}{0cm}\centering\hsize=\linewidth
\begin{tikzpicture}[scale=0.95,text depth=-0.2ex,text height=1.2ex]
\draw(2.3,0.9) node {\footnotesize Merged copy $\FSa_1$};
\Dnine\D\D\D\D\E{\f_6}{\f_7}{\f_8}{\f_9}
\begin{scope}[xshift=6cm]
\draw(2.3,0.9) node {\footnotesize Merged copy $\FSa_2$};
\Dnine\D\D\D\E\E{\f_6}{\f_7}{\f_8}\E
\end{scope}
\begin{scope}[xshift=12cm]
\draw(2.3,0.9) node {\footnotesize Merged copy $\FSa_3$};
\Dnine\D\E\E\E\E{\f_6}\E\E\E
\end{scope}
\end{tikzpicture}%
\caption{Three possible results of synchronizing the filesystems in Figure
\ref{fig:update}}\label{fig:mergers}
\end{adjustwidth}\end{figure}
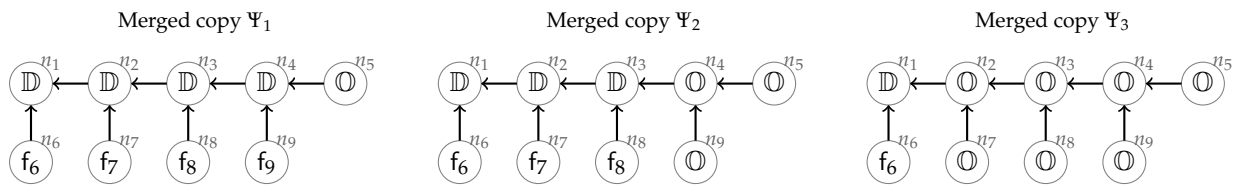

Theorem \ref{thm:merge} is illustrated on the filesystems in Figure
\ref{fig:update}. The canonical sets transforming the original filesystem to
the replicas are $A=\{\sigma_1, \sigma_2, \sigma_3, \sigma_4, \sigma_5\}$
(where $\sigma_i = \(n_i,\D,\E)$) and $B=\{\tau_5, \tau_6, \tau_7, \tau_8,
\tau_9\}$ (where $\tau_i=\(n_i,\E\D,\f_i)$), respectively. The synchronized
states shown on Figure \ref{fig:mergers} correspond to the mergers $M_1 = \{
\sigma_5, \tau_6, \tau_7, \tau_8, \tau_9 \}$, $M_2= \{ \sigma_4, \sigma_5,
\tau_6, \tau_7, \tau_8 \}$, and $M_3 = \{ \sigma_2, \sigma_3, \sigma_4,
\sigma_5, \tau_6 \}$, respectively. Altogether there are six mergers of $A$
and $B$ depending on how many levels of the directory erasures in $A$ are
kept. To get $\FSa_1$ from the replica $\FS_1$ we first roll back the
commands $\sigma_1$, $\sigma_2$, $\sigma_3$ and $\sigma_4$ in this order
(commands in the set $A\setm M_1$), which restores the four directories.
Then we apply the commands $\tau_6$, $\tau_7$, $\tau_8$, $\tau_9$ from the
other command set $B$ to set the file values. Getting $\FSa_1$ from the
second replica requires fewer commands. First roll back $\tau_5$ (the only
member of $B\setm M_1$) restoring the directory at $n_5$, then apply
$\sigma_5$ to erase this directory.

\medskip

The proof of Theorem \ref{thm:merge} uses some intricate properties of 
canonical sets and mergers which we prove first.

\begin{Proposition}\label{prop:meet}
Let $A$ and $B$ be refluent canonical sets with $\tau\in A$, $\sigma \in
A\cup B$ and $\sigma\orel \tau$. Then $\sigma \in A$.
\end{Proposition}
\begin{proof}
Let $\sigma$, $\tau$ be on nodes $n$ and $m$, respectively, and $\FS$ be any
filesystem on which both $A$ and $B$ work. The value $\FS(n)$ is the input
value of $\sigma$, and $\FS(m)$ is the input value of $\tau$. Now $\sigma
\orel \tau$ matches either the pattern $\(n,\D\F,\E) \orel \(\parent
n,\D,\F\E)$ or the pattern $\(\parent m,\E\F,\D) \orel \(m,\E,\F\D)$, see
Definition \ref{def:ex-order}. If $A$ has no command on node $n$, then
executing $\tau$ would break $\FS$ as $\FS(n)$ still has its original value.
Thus let $\sigma'\in A$ be on node $n$. The input value of $\sigma'$ is the
same as that of $\sigma$, moreover $\sigma' \orel \tau$ (as $A$ is canonical
and $\sigma'$ and $\tau$ are on parent--child nodes). Then the output values
of $\sigma'$ and $\sigma$ are also equal since they are either both $\E$ or
$\D$, thus $\sigma=\sigma'$.
\end{proof}

A similar statement holds for mergers, but the proof is more involved.

\begin{Proposition}\label{prop:toM}
Let $A$ and $B$ be refluent canonical sets, $M\subseteq A\cup B$ be a merger
with $\tau\in M$, $\sigma \in A\cup B$ and $\sigma\orel \tau$. Then $\sigma
\in M$.
\end{Proposition}
\begin{proof}
Let $\sigma$ and $\tau$ be a lowest counterexample to the claim, and $n$ and
$m$ be their respective nodes. By ``lowest'' we mean that there is no other
counterexample pair $\sigma^*, \tau^*$ where the node of $\sigma^*$ would be
below $n$.

The argument of the proof of Proposition \ref{prop:meet} gives that if $M$
has a command on $n$ (the node of $\sigma$), then $\sigma\in M$. Thus
$\sigma$ cannot be added to $M$ only if $M$ already contains a command
$\sigma'$ on a node $n'$ such that $n'$ and $n$ are comparable and $\sigma'$
and $\sigma$ cannot be connected by an $\orel$-chain. If $m$ and $n'$ are
comparable, then there is an $\orel$-chain in $M$ between $\tau$ and
$\sigma'$ as both are in $M$. As no command in $M$ is on $n$, the node $n$
cannot be between $m$ and $n'$. But then $\sigma \orel \tau \orel \cdots
\orel \sigma'$ is an $\orel$-chain in $M$ connecting $\sigma$ and $\sigma'$,
a contradiction. (Note that the chain between $\tau$ and $\sigma'$ cannot be
in the other direction.)

Therefore $m$ and $n'$ are uncomparable, which means that $m$ and $n'$ are
both below $n$, consequently $\sigma$ and $\tau$ are constructors. Let $\FS$
be any filesystem on which both $A$ and $B$ work. As $\sigma$ matches
$\(n,\E\F,\D)$, $\FS(n)$ is either empty or is a file. It means that all
nodes below $n$ are empty, in particular $\FS(n')=\E$, and $\sigma'$ changes
this empty value to something else. Now suppose $\sigma'\in A$. As $A$ does
not break $\FS$, it must change the non-directory value at $n$ to a
directory, consequently $\sigma\in A$ as well. Then there is an
$\orel$-chain between $\sigma$ and $\sigma'$ in A starting with $\sigma$ and
ending with $\sigma'\in M$. As $\sigma$ was chosen to be the lowest
counterexample, all elements of this chain are in $M$, which is a
contradiction again.
\end{proof}

Recall from Proposition \ref{prop:5} that $B \subc A$ if and only if
$B\subseteq A$ and there are no $\sigma\orel\tau$ such that $\tau\in B$ and
$\sigma \in A\setm B$.

\begin{Proposition}\label{prop:thm-bc}
Let $A$, $B$ be refluent canonical sets, and $M$ be a merger. Then $A\cap M
\subc A$, and $A\cap M \subc M$.
\end{Proposition}

\begin{proof}
Both statements follow from the combination of Propositions \ref{prop:meet}
and \ref{prop:toM}. If $\tau\in A \cap M$ and $\sigma \orel \tau$ with
$\sigma\in A\cup B$, then $\sigma \in A\cap M$.
\end{proof}

Claim (a) of Theorem \ref{thm:merge} follows immediately from Proposition
\ref{prop:thm-bc}. Let $A_2 = A\cap M$, $A_1=A\setm M$, and $B'=M\setm A$.
By Propositions \ref{prop:thm-bc} and \ref{prop:5} we have $A=A_2 \circ
A_1$, and $M=A_2 \circ B'$. Thus
\begin{disp}
  M = A_2\circ B' \semge (A_2\circ A_1 \circ A_1^{-1}) \circ B'
     = A \circ A_1^{-1} \circ B',
\end{disp}
and a similar reasoning gives claim (b).

\medskip

Recall that the notation $\sigma \indep \tau$ means that the nodes which the
commands $\sigma$ and $\tau$ are acting on are uncomparable (that is,
independent). We write $\sigma \indep B$ to mean that for every $\tau\in B$
we have $\sigma \indep \tau$, and $A \indep B$ to mean that for every
$\sigma\in A$ we have $\sigma \indep B$.

\begin{Proposition}\label{prop:indepAB}
Suppose that the canonical sets $A$ and $B$ are applicable to the filesystem
$\FS$; moreover, $A\setm B \indep B\setm A$. Then $A\cup B$ is canonical and is
applicable to $\FS$.
\end{Proposition}
\begin{proof}
First we show that $A\cup B$ is canonical. This is so as $A\setm B$ and
$B\setm A$ do not have commands on comparable or coinciding nodes, thus any
two commands in $A\cup B$ on comparable nodes are both in $A$, or are both
in $B$. Second, $A\cup B$ is applicable to $\FS$ when $A$ and $B$ are
disjoint as in this case, by assumption, $A\indep B$, see also \cite[Lemma
36]{Csi16}.

When $A$ and $B$ are not disjoint, let $C=A \cap B$. By Proposition
\ref{prop:meet}, $C \subc A$, thus $A \equiv C\circ (A\setm C)$ which means
$A\FS = (A\setm C)(C\FS)$. Similarly, $B\FS = (B\setm C)(C\FS)$. Applying
the first case of this proof to the disjoint sets $A\setm C$, $B\setm C$ and
the filesystem $C\FS$ shows that $(A\cup B)\setm C$ is applicable to $C\FS$.
Since $C \subc (A\cup B)$ also holds, we get that $A\cup B$ is applicable to
$\FS$.
\end{proof}

After these preparations claim (c) of Theorem \ref{thm:merge} can be proved
as follows. Let $A_2=A\cap M$ and $B_2=B\cap M$. Then $A_2\cup B_2=M$ and
$M\setm A_2= B_2\setm A_2$. Now $A_2\subc M$ by Proposition
\ref{prop:thm-bc}, which means that $A_2$ is canonical and $M\equiv A_2\circ
(M\setm A_2) = A_2\circ(B_2\setm A_2)$, and similarly $M\equiv B_2\circ
(A_2\setm B_2)$. We claim that $A_2\setm B_2\indep B_2\setm A_2$. This is
because these sets are disjoint and if $\sigma$ in the first set and $\tau$
in the second set are on comparable nodes, then they are $\orel$-connected
(as both are in the canonical set $M$), thus in any ordering of $M$ which
honors $\orel$ they are in a fixed order. But in $A_2\circ (B_2\setm A_2)$
$\sigma$ comes before $\tau$, and in $B_2\circ (A_2\setm B_2)$ $\tau$ comes
before $\sigma$, which is a contradiction.

Let $\FS$ be a filesystem to which both $A$ and $B$ can be applied. Again by
Proposition \ref{prop:thm-bc}, $A_2\subc A$, thus $A_2$ is also applicable
to $\FS$, and similarly $B_2$ is applicable to $\FS$. By Proposition
\ref{prop:indepAB}, $M=A_2\cup B_2$ is also applicable to $\FS$, which is
the first claim in Theorem \ref{thm:merge} (c). For the other two claims
observe that $A_1=A\setm M=A\setm A_2$ and $B'=M\setm A=M\setm A_2$. We know
$A_2\subc A$, thus $A\circ A_1^{-1} =A_2\circ A_1\circ A_1^{-1}$, which is
applicable to $\FS$ and gives $A_2\FS$. Since $A_2\subc M$ and $M$ is
applicable to $\FS$, $M\setm A_2=B'$ is applicable to $A_2\FS$. All
together, $A\circ A_1^{-1}\circ B'$ is applicable to $\FS$, as claimed.
Similar reasoning gives that $B \circ B_1^{-1} \circ A'$ is also applicable
to $\FS$.

\section{Synchronization by conflict resolution}\label{sec:conflict}

In the realm of file synchronization a \emph{conflict} is represented by two
filesystem commands which can be executed individually, but not together.
Resolving the conflict means choosing one of the following actions (see,
among others, \cite{syncpal-thesis,NS16}):
\begin{itemz}
\item discard both commands,
\item discard one command and keep the other (loser/winner paradigm), or
\item replace one or both commands by a new one.
\end{itemz}
Only the second alternative satisfies both the \emph{intention-confined
effect} principle \hyperlink{R1}{R1} (use only commands issued by the user), and the
\emph{aggressive effect preservation} principle \hyperlink{R2}{R2} (carry over as many
operations as possible). Our synchronizer uses this loser/winner paradigm to
solve all conflicts. Resolving a conflict might create new ones, or
automatically solve -- or make irrelevant -- others. Thus after each step
the set of conflicts has to be regenerated, or at least
revalidated/reviewed. Termination of the synchronization process is not
automatic and has to be proved, see \cite{syncpal-thesis}. Fortunately, in
our case no new conflicts are generated, and resolved and disappearing
conflicts can be identified easily. As each step eliminates at least one
conflict, the synchronizer clearly stops after finitely many steps.

\subsection{Theoretical foundation}\label{subsec:foundation}

In this subsection we fix the refluent canonical sets $A$ and $B$ and the
merger $M$ which the synchronization is supposed to reach, but is not
necessarily known at the start. First we show that synchronization can be
reduced to the case when $A$ and $B$ are disjoint (though they might contain
different commands on the same node). Then we define when a command pair
represents a \emph{conflict}, and show that this notion has the expected
properties: a merger $M$ does not contain conflicting pairs, and if there
are no conflicts, then $M$ is the disjoint union of $A$ and $B$. The main
novelty of our conflict resolution algorithm is that instead using the
winner to build up the merger, it only discards the losing command from
future considerations. In other words, the algorithm thins out the sets $A$
and $B$ until only the merger $M$ remains.

\begin{Proposition}\label{prop:intersectionM}
Let $A, B$ be refluent canonical sets, and $M$ be a merger. Then
(a) $A\cap B\subc A$ and $A\cap B\subc B$,
(b) $A\cap B\subseteq M$, and
(c) $A\cap B\subc M$.
\end{Proposition}
\begin{proof}
Claim (a) follows from Proposition \ref{prop:meet} since if $\tau\in A\cap
B$, $\sigma\in A\cup B$ and $\sigma\orel\tau$, then $\sigma\in A\cap B$.
From this and from (b) claim (c) follows, too.

To prove (b) it suffices to check that $M\cup(A\cap B)$ is canonical. First,
all commands in this union are on different nodes. Both $M$ and $A\cap B$
are canonical, so if their union were not, then there are commands
$\sigma\in A\cap B$ and $\sigma'\in M\setm(A\cap B)$ on comparable nodes
which are not connected by an $\orel$-chain in $M\cup(A\cap B)$. Assume
$\sigma'\in A$. There is a chain between $\sigma$ and $\sigma'$ in $A$.
Since $A\cap B \subc A$, we know this chain has the direction $\sigma \orel
\cdots \orel \sigma'$. From this and because $\sigma'\in M$, we know all
members of this chain are in $M$ based on Proposition \ref{prop:toM}, which
is a contradiction.
\end{proof}

Let $C=A\cap B$ and let $\FS$ be a filesystem which shows that $A$ and $B$
are refluent, that is, both $A$ and $B$ work on $\FS$. Let $A'=A\setm C$,
$B'=B\setm C$, and $M'=M\setm C$; observe that the command sets $A'$ and
$B'$ are disjoint. By Proposition \ref{prop:intersectionM} $C$ is canonical
and can be moved to the front of $A$ and $B$ and $M$; moreover $A'$ and $B'$
and $M'$ are also canonical. Now $A'$ and $B'$ are refluent (both are
applicable to $C\FS$), and $M'$ is a merger for $A'$ and $B'$. The converse
of the last statement is also true: if $M^*$ is any merger of $A'$ and $B'$,
then $C\cup M^*$ is a merger of $A$ and $B$. Consequently it suffices to
find the merger $M'$ of the disjoint command sets $A'$ and $B'$, and then
adding the intersection $A\cap B$ to $M'$. For this reason, from this point
on we assume that $A$ and $B$ share no commands.

\begin{Definition}\label{def:conflict}
Let $A$ and $B$ be fixed disjoint, refluent canonical sets. The pair
$(\sigma,\tau)$ is a \emph{conflict} if $\sigma\in A$, $\tau\in B$ and
$\sigma\notindep \tau$, that is, the nodes of $\sigma$ and $\tau$ are
comparable (including when they coincide).
\end{Definition}

\begin{figure}[H]\begin{adjustwidth}{-\extralength}{0cm}\centering\hsize=\linewidth
\begin{tikzpicture}
\begin{scope}[xshift=-1cm]
\foreach\x in {1,...,5}{
    \coordinate (x\x) at (1.1*\x,-0.2);
    \coordinate (y\x) at (1.1*\x,-0.9);
}
\fill[color=red!10!white] (x2)+(0,0.3) circle(0.25cm);
\fill[color=green!10!white] (y2)+(0,-0.3) circle(0.25cm);
\foreach\x in {1,...,5}{
    \draw (x\x)+(0,0.3) node {$\sigma_\x$};
}
\foreach \x in{6,7,8,9}{
    \draw (1.1*\x-5.5,-1.2) node {$\tau_\x$};
}
\draw (5.5,-1.2) node {$\tau_5$};
\foreach \x in {1,...,5}{\draw (x1)--(y\x);}
\foreach \x in {3,4,5}{\draw (x2)--(y\x);}
\draw (x3)--(y3)  (x4)--(y4) (x5)--(y5)--(x3)--(y4) (x4)--(y5);
\draw (3.3,-1.7) node{\footnotesize a)};
\draw[line width=2pt] (x2)--(y2);
\end{scope}

\begin{scope}[xshift=5.5cm]
\foreach\x in {1,...,5}{
    \coordinate (x\x) at (1.1*\x,-0.2);
    \coordinate (y\x) at (1.1*\x,-0.9);
}
\fill[color=black!17!white](x1)+(0,0.3) circle(0.3cm)
                          (x2)+(0,0.3) circle(0.3cm);
\fill[color=green!10!white] (x4)+(0,0.3) circle(0.25cm);
\fill[color=red!10!white] (y5)+(0,-0.3) circle(0.25cm);

\foreach\x in {1,...,5}{
    \draw (x\x)+(0,0.3) node {$\sigma_\x$};
}
\foreach \x in{6,7,8,9}{
    \draw (1.1*\x-5.5,-1.2) node {$\tau_\x$};
}
\draw (5.5,-1.2) node {$\tau_5$};
\draw (y3)--(x3)--(y4)--(x4) (y5)--(x5) (x3)--(y5);
\draw[line width=2pt] (x4)--(y5);
\draw (3.3,-1.7) node{\footnotesize b)};
\end{scope}

\begin{scope}[xshift=12cm]
\foreach\x in {1,...,5}{
    \coordinate (x\x) at (1.1*\x,-0.2);
    \coordinate (y\x) at (1.1*\x,-0.9);
}
\fill[color=black!17!white](x1)+(0,0.3) circle(0.3cm)
                          (x2)+(0,0.3) circle(0.3cm)
                          (y4)+(0,-0.3) circle(0.3cm)
                          (y5)+(0,-0.3) circle(0.3cm);
\fill[color=green!10!white] (y3)+(0,-0.3) circle(0.25cm);
\fill[color=red!10!white] (x3)+(0,0.3) circle(0.25cm);

\foreach\x in {1,...,5}{
    \draw (x\x)+(0,0.3) node {$\sigma_\x$};
}
\foreach \x in{6,7,8,9}{
    \draw (1.1*\x-5.5,-1.2) node {$\tau_\x$};
}
\draw (5.5,-1.2) node {$\tau_5$};
\draw[line width=2pt] (x3)--(y3);

\draw (3.3,-1.7) node{\footnotesize c)};
\end{scope}

\end{tikzpicture}%
\caption{\fontsize{9}{10.2}\selectfont
a) The conflict graph of the synchronization problem of Figure
\ref{fig:update}. The conflict $(\sigma_2,\tau_7)$ is resolved with
the winner $\tau_7$.\\
b) The conflict graph after resolving $(\sigma_2,\tau_7)$;
commands $\sigma_1$ and $\sigma_2$ are deleted. The next conflict we
resolve is $(\sigma_4,\tau_5)$.\\
c) The conflict graph after resolving $(\sigma_4,\tau_5)$ with the winner
$\sigma_4$; commands $\tau_9$, $\tau_5$ are deleted. The final conflict
graph contains the commands $\sigma_4$, $\sigma_5$, $\tau_6$, $\tau_7$,
$\tau_8$, and no edges.
}\label{fig:conflict-graph}
\end{adjustwidth}\end{figure}
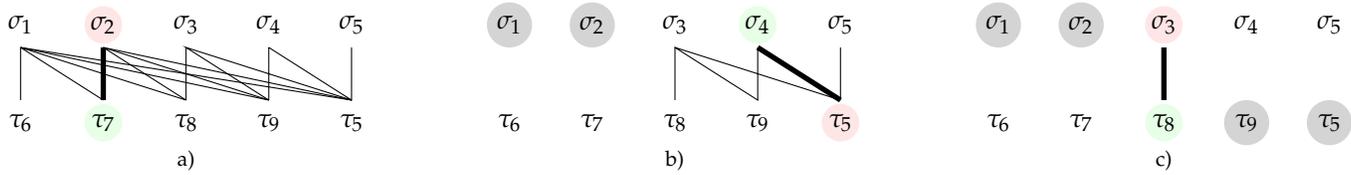

The \emph{conflict graph} is the bipartite graph with the disjoint command
sets $A$ and $B$ as the two classes, and an edge between $\sigma\in A$ and
$\tau\in B$ if they are in conflict. Figure \ref{fig:conflict-graph} a)
shows the conflict graph of the synchronization problem depicted on Figure
\ref{fig:update}. As an example, command $\sigma_5$ is connected to $\tau_5$
only, as both $\sigma_5$ and $\tau_5$ are on node $n_5$, while $n_5$ is
independent of the nodes $n_6$, \dots, $n_9$ on which the other commands in
$B$ are.

\begin{Proposition}\label{prop:conflict-AB}
It is impossible that 
both commands of the conflict $(\sigma,\tau)$ are in $M$.
\end{Proposition}
\begin{proof}
The meanings of the symbols are as in Definition \ref{def:conflict}. Suppose
by contradiction that both are in $M$. As they are on comparable nodes,
there is an $\orel$-chain in $M$ between $\sigma$ and $\tau$. Consequently
there are two consecutive commands in this chain so that one is in $A$ and
the other is in $B$, say $\sigma'\orel\tau'$, $\sigma'\in A$ and $\tau'\in
B$. But Proposition \ref{prop:meet} says that in this case $\sigma'\in B$,
contradicting that $A$ and $B$ are disjoint.
\end{proof}

\begin{Proposition}\label{prop:orel}
Suppose $(\sigma,\tau)$ is a conflict, $\sigma'\in A$ and $\sigma\orel
\sigma'$. Then $(\sigma',\tau)$ is also a conflict.
\end{Proposition}
\begin{proof}
Let $\sigma$, $\sigma'$ be on nodes $n$ and $n'$, respectively. If $n'$ is
the parent of $n$, then the node of $\tau$ is comparable to $n'$, thus we
are done. Therefore $n$ is the parent of $n'$, $\sigma$ and $\sigma'$ are
constructors, and $\sigma$ matches $\(n,\E\F,\D)$. If the node of $\tau$ is
not below $n$, then we are done. If it is, then the input type of $\tau$ is
$\E$ (as in the original filesystem $\FS$, $\FS(n)$ is not a directory,
therefore every node below it is empty). For the same reason the command set
$B$ must contain a command on node $n$ (otherwise when executing $\tau$,
$\FS$ still has the original content at $n$, and then $\tau$ breaks $\FS$).
This command has the same input type as $\sigma$, must create a directory,
thus it is equal to $\sigma$, contradicting that $A$ and $B$ are disjoint.
\end{proof}

By symmetry, the same claim holds when $A$ and $B$ are swapped. Fix
$\sigma\in A$, and split $B$ into $B^\no_\sigma$ and $B^\yes_\sigma$ as
follows:
\begin{disp}\begin{array}{r@{\;=\;}l}
  B^\no_\sigma & \{ \tau\in B : \mbox{ $(\sigma,\tau)$ \emph{is not} a conflict} \}
   = \{ \tau\in B: \sigma \indep \tau \}, \\[4pt]
  B^\yes_\sigma & \{ \tau\in B : \mbox{ $(\sigma,\tau)$ \emph{is} a conflict} \}
   = \{ \tau \in B: \sigma\notindep \tau \}.
\end{array}\end{disp}

We know $B^\no_\sigma \subc B$ by Propositions \ref{prop:5} and
\ref{prop:orel}, which means that if $\FS$ is a filesystem $B$ does not
break, then neither does $B^\no_\sigma$. Also, if $\sigma \orel \sigma'$,
then $B^\yes_\sigma \subseteq B^\yes_{\sigma'}$. In particular, if
$B^\yes_{\sigma'}$ is empty, then so is $B^\yes_\sigma$.
Next we show that if there are no conflicts, then the synchronization is
done; however, we need a stronger statement which will be used to justify
the correctness of the synchronization process.

\begin{Proposition}\label{prop:indep}
Suppose $B^\yes_\sigma$ is empty, that is, no $\tau\in B$ is in conflict with
$\sigma\in A$. Then $\sigma\in M$.
\end{Proposition}
\begin{proof}

It is enough to show that $M\cup \{\sigma\}$ is canonical. Since
$\sigma\indep B$, it has at most one command on every node. It can only be
non-canonical if there is a $\tau\in M$ on a related node that cannot be
connected to $\sigma$ using an $\orel$-chain from the commands in $M$. This
$\tau$ must be in $A$, which has a chain between $\sigma$ and $\tau$. If
this chain goes in the direction $\sigma \orel \cdots \orel \tau$, then
$\sigma\in M$ by Proposition \ref{prop:toM}. Otherwise we know $\tau\orel
\cdots \orel \sigma$, and based on the remark above, $B^\yes_{\omega}$ is
empty for all members of this chain and so are independent of $B$. This
means the chain can be added to $M$, which is a contradiction.
\end{proof}

\begin{Corollary}[Termination]\label{corr:termination}
If there are no conflicts, then the merger $M$ is $A \cup B$.
\end{Corollary}
\begin{proof}
By Proposition \ref{prop:indep} every command in $A$ is in $M$, and
similarly $B\subseteq M$. As $M$ is a subset of $A\cup B$, they must be
equal.
\end{proof}

\subsection{The synchronization algorithm}

The input of the synchronization algorithm is $(A^o$, $B^o)$,
which is
a pair of
refluent canonical command sets. The goal is to find, using successive
conflict resolution steps, a merger command set $M$ that fits Definition
\ref{def:merger}. 
In addition,
we aim to show that our algorithm can produce
\emph{any} of the mergers allowed by the definition, and only valid mergers.
To achieve this, the algorithm accepts a pre-set merger $M^o$ as an optional
argument. Whenever the algorithm needs to make an otherwise arbitrary
choice, $M^o$ is used to choose one of the possibilities. While choosing any
of the possibilities will generate a valid merger, guiding the choice will
guarantee that $M^o$ will be generated.

\begin{figure}[H]
\centering
\begin{tikzpicture}[shorten >=2pt,yscale=0.7]
\node[draw] (q0) at (1.5,3) {$\scriptstyle C=A^o\cap B^o$};
\node[draw] (q1) at (3.5,2) {$\scriptstyle A^o\setm C$};
\node[draw] (q2) at (3.5,1) {$\scriptstyle B^o\setm C$};
\node[draw] (q3) at (3.5,0) {$\scriptstyle M^o\setm C$};
\node[draw] (r1) at (8,1) {$\scriptstyle M'\cup C$};
\coordinate (qq1) at (-0.4,2);
\coordinate (qq2) at (-0.4,1);
\coordinate (qq3) at (-0.4,0);
\draw (qq1) node {$\scriptstyle A^o$};
\draw (qq2) node {$\scriptstyle B^o$};
\draw (qq3) node {$\scriptstyle M^o$};
\draw (9.2,1) node {$\scriptstyle M$};
\draw (6.8,1) node {$\scriptstyle M'$};
\path[->] (qq1)+(0.3,0) edge (q1) (qq2)+(0.3,0) edge (q2) (qq3)+(0.3,0) edge (q3);
\path[->] (1.5,1) edge (q0);
\draw (0.3,2)-- (0.3,3.1);
\path[->] (0.3,3) edge (q0);
\path (q0) edge (8.1,3);
\path[->] (7.05,1) edge (r1) (r1) edge (9,1) (8,3) edge (r1);
\draw (5,2.2) rectangle (7,-0.2);
\path[->] (q1) edge (4.96,2) (q2) edge (4.96,1) (q3) edge (4.96,0);
\draw (5.8,1.4) node {call};
\draw (5.8,0.7) node {\textbf{Alg. 2}};
\draw (2.5,3)--(2.5,0.1);
\draw[->] (2.5,2.2) -- (3.0,2.2);
\draw[->] (2.5,1.2) -- (3.0,1.2);
\draw[->] (2.5,0.2) -- (2.96,0.2);
\end{tikzpicture}%
\caption{\fontsize{9}{10}\selectfont
Outline of Algorithm 1. Given the input $(A^o, B^o, M^o)$, extract the common
part $C$ of $A^o$ and $B^o$, and call Algorithm 2 with the reduced sets.
Finally, add $C$ to the returned $M'$ to generate the output $M$.%
}\label{fig:alg1}
\end{figure}
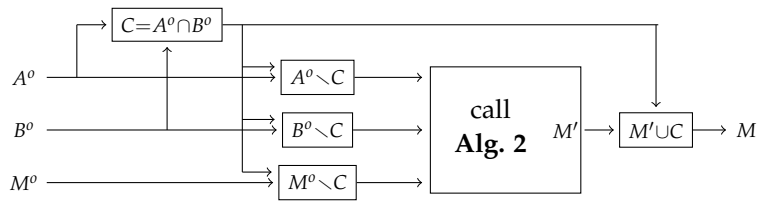

\begin{figure}[H]\begin{adjustwidth}{-\extralength}{0cm}\centering\hsize=\linewidth
\centering
\begin{tikzpicture}[yscale=0.7]
\def\ss{\scriptstyle}
\coordinate (qq1) at (-0.25,1.9);
\coordinate (qq2) at (-0.25,1.1);
\coordinate (qq3) at (-0.25,0.1);
\draw[->] (0,1.9)--(0.35,1.9);
\draw[->] (0,1.1)--(0.35,1.1);
\draw (qq1) node {$\ss A^o$};
\draw (qq2) node {$\ss B^o$};
\draw (qq3) node {$\ss M^o$};
\draw (0.4,2.2) rectangle (1.6,0.8);
\draw (1,1.8) node {$\ss A^0\to A$} (1,1.2) node {$\ss B^0\to B$};
\draw (1,2.8) node {\footnotesize\it Initialize:};

\draw[->] (1.6,1.5) -- (2.9,1.5);

\draw (3,2.2) rectangle (5,0.8);
\draw (4,1.8) node{\footnotesize find conflict};
\draw (4,1.2) node{$\ss \sigma\in A,~ \tau\in B$};

\draw[->] (5,1.5)--(5.9,1.5);
\draw (5.42,1.65) node {\footnotesize found};
\draw (4,0.8)--(4,0.4)--(12,0.4)--(12,1.5);
\draw[->] (12,1.5)--(12.45,1.5);
\draw (3.38,0.50) node {\footnotesize not found};
\draw (0,0.1)--(7,0.1); \draw[->] (7,0.1)--(7,0.7);

\draw (6,2.2) rectangle (8,0.8);
\draw (7,1.8) node {\footnotesize determine};
\draw (7,1.2) node {\footnotesize the winner};

\draw[->] (8,1.5)--(8.9,1.5);

\draw (9,2.2) rectangle (11,0.8);
\draw (10,1.8) node {\footnotesize reduce set};
\draw (10,1.2) node {\footnotesize of loser};

\draw (11,1.5)--(11.3,1.5)--(11.3,2.6)--(4,2.6);
\draw[->] (4,2.6)--(4,2.25);

\draw (7,3.0) node {\footnotesize\it Iterate:};

\draw (12.5,2.2) rectangle (14.5,0.8);
\draw (13.5,1.5) node {$\ss M=A\cup B$};
\draw[->] (14.5,1.5) -- (15,1.5);
\draw (15.2,1.5) node{$\ss M$};

\draw (13.5,2.8) node {\footnotesize\it Finish:};

\end{tikzpicture}%
\caption{\fontsize{9}{10}\selectfont
Outline of Algorithm 2. The \emph{Initialize} step sets $A$ and $B$. The
\emph{Iterate} step is executed until no more conflicts are found. Conflicts
are
resolved using the hint from the additional input $M^o$, then the command
sets $A$ and $B$ are reduced. If no conflict remains, the merger $M=A\cup B$
is returned by the \emph{Finalize} step.%
}\label{fig:alg2}
\end{adjustwidth}
\end{figure}
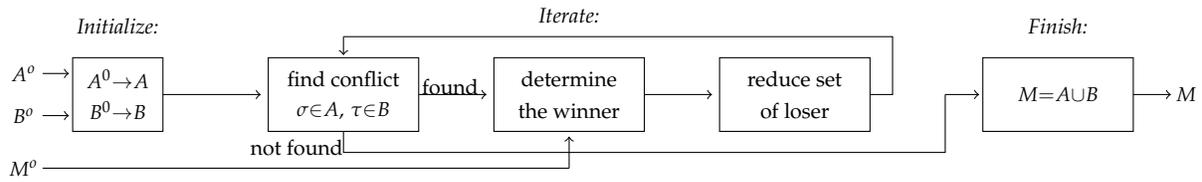

According to the discussion in Section \ref{subsec:foundation}, the
synchronization algorithm has two main components. The first one, called
Algorithm \ref{alg:1}, reduces the general problem to the special case when
the input command sets have no common elements. Properties of the merger set
proved in Proposition \ref{prop:intersectionM} provide the foundation of the
reduction.
The second component, called Algorithm \ref{alg:2}, assumes that its input 
command sets $A^o$ and $B^o$ are disjoint so that it can rely on the
intrinsic properties of the conflict graph when generating the merger $M$.
The two algorithms are visalized in Figures \ref{fig:alg1} and \ref{fig:alg2},
respectively. The detailed description is followed by the proofs of the 
correctness and a short discussion on time and space complexity.

\begin{Algorithm}[Synchronization in the general case]\label{alg:1}\upshape
Let $C=A^o \cap B^o$. Execute Algorithm \ref{alg:2} with the command sets
$A^o\setm C$ and $B^o\setm C$. If $M^o$ is specified, then it must satisfy
$C\subseteq M^o$; in this case pass $M^o\setm C$ as the optional input to 
Algorithm \ref{alg:2}.
When Algorithm \ref{alg:2} returns $M'$, return $M=C\cup M'$.
\end{Algorithm}
\begin{proof}[Correctness of Algorithm \ref{alg:1}]
As discussed after the proof of Proposition \ref{prop:intersectionM},
$A^o\setm C$ and $B^o\setm C$ are refluent canonical sets, and 
$C$ is an initial segment of every merger of $A^o$ and $B^o$. Moreover, if
$M^o$ is a merger of $A^o$ and $B^o$, then $M^o\setm C$ is a merger of
$A^o\setm C$ and $B^o\setm C$.
From here the correctness of the algorithm follows easily.
\end{proof}

\begin{proof}[Complexity of Algorithm \ref{alg:1}]
Apart from the time and space requirements of Algorithm 2, the five
operations here require computing the union, intersection and difference of sets.
Using hashing techniques, these operations can be performed both in time and space
linear in the input size.
\end{proof}

Inputs of the main Algorithm \ref{alg:2} are the \emph{disjoint} refluent
canonical sets $A^o$ and $B^o$, and the optional target merger $M^o$ of
$A^o$ and $B^o$. The goal is to produce a merger $M$. If $M^o$ is given,
then, at the end, $M$ and $M^o$ must be equal.

The algorithm maintains two command sets $A$ and $B$, initially set
to $A^o$ and $B^o$. During execution these sets satisfy the
following invariants:
\begin{itemz}
\item[1.] $A\subc A^o$, $B\subc B^o$, consequently $A$ and $B$ are refluent
canonical subsets of $A^o$ and $B^o$;
\item[2.] all mergers of $A$ and $B$ are mergers of $A^o$ and $B^o$;
\item[3.] if $M^o$ is specified, then $M^o$ is a merger of $A$ and $B$.
\end{itemz}
In each iteration one or more commands are deleted either from $A$ or from
$B$ until only the commands of the merger remain.

\begin{Algorithm}[Synchronization of disjunct sets]\label{alg:2}\upshape
\textit{Initialize:\space} Set $A^o\to A$ and $B^o\to B$.
\begin{list}{}{\setlength\labelwidth{0pt}\setlength\itemindent{-\leftmargin}\setlength\itemsep{0pt}\setlength\parsep{2pt
plus 1pt minus 1pt}\setlength\topsep{2pt plus 2pt minus 1pt}}
\item\textit{Iterate:\space} Choose $\sigma\in A$ and $\tau\in B$ such that
$(\sigma,\tau)$ is a conflict. If no such pair exists, go to Finish.
If $M^o$ is given, choose a conflict so that either $\sigma$ or
$\tau$ is in $M^o$.

\textit{Conflict resolution:\space} Out of $\sigma$ and $\tau$, choose the
\emph{winner} and the \emph{loser} commands. If $M^o$ is given, let the
winner be the one in $M^o$. If $\sigma$ is the winner, delete all commands
from $B$ which are in conflict with $\sigma$ (including $\tau$). If $\tau$
is the winner, delete all commands from $A$ which are in conflict with
$\tau$. Go back to Iterate.
\item\textit{Finish:\space} Return $A\cup B$ as the merger.
\end{list}
\end{Algorithm}

Execution of Algorithm \ref{alg:2} is illustrated in Figure
\ref{fig:conflict-graph}. The initial command sets are $A=\{\sigma_1, \dots,
\sigma_5\}$ and $B=\{\tau_6, \dots, \tau_5\}$. In the first iteration
$\tau_7 \in B$ is the winner; therefore, commands in $A$ which are in
conflict with $\tau_7$ ($\sigma_1$ and $\sigma_2$) are removed from $A$. In
the next iteration the winner is $\sigma_4\in A$, and $\tau_9$ and $\tau_5$
are removed from $B$. In the last iteration $\sigma_3$ is removed from $A$,
thus the remaining sets are $A=\{\sigma_4,\sigma_5\}$ and
$B=\{\tau_6,\tau_7,\tau_8\}$. The returned merger is $A\cup B$.

\begin{proof}[Correctness of Algorithm \ref{alg:2}]
After the initialization step all three invariants hold. The algorithm
terminates as in each iteration the total number of remaining conflicts
strictly decreases. When it terminates, the returned set $A\cup B=M$ is a
merger of the original command sets $A^o$ and $B^o$. This is because at that
point there are no conflicts between $A$ and $B$, thus, by Corollary
\ref{corr:termination}, $M$ is a merger of $A$ and $B$---in fact, the only
one---, and then the second invariant guarantees that $M$ is also a merger
of $A^o$ and $B^o$. If $M^o$ was submitted to the algorithm, then the third
invariant ensures $M=M^o$.
Thus we only need to show that the iteration preserves all three invariants.
Suppose there is a conflict $(\sigma,\tau)$ between $A$ and $B$, and the
winner of the conflict is $\sigma\in A$.

\textit{Invariant 1.} Commands in conflict with $\sigma$ are deleted from
$B$, thus $B$ is replaced by the set
\begin{disp}
B^\no_\sigma=\{ \tau'\in B: \mbox{ $(\sigma,\tau')$ is not a conflict}\}.
\end{disp}
Proposition \ref{prop:orel} (see the discussion following the proof) and the
invariant imply $B^\no_\sigma\subc B\subc B^o$. Consequently the invariant
$B\subc B^o$ remains true.

\textit{Invariant 2.}
We know that if a merger of $A$ and $B$ contains $\sigma$, then it cannot
contain any of the discarded elements of $B$ (see Proposition
\ref{prop:conflict-AB}). We also know that all mergers of $A$ and
$B^\no_\sigma$ contain $\sigma$ by Proposition \ref{prop:indep} as it is not
in conflict with any other command. Consequently the set of mergers of $A$
and $B^\no_\sigma$ consists of those mergers of $A$ and $B$ that contain
$\sigma$, and the invariant continues to hold.

\textit{Invariant 3.}
If $M^o$ was specified for the algorithm, then we chose the winner
accordingly, and so $\sigma\in M^o$. A consequence of the discussion above
is that $M^o$ remains part of the set of mergers of $A$ and $B^\no_\sigma$.

Finally we show that if $M^o$ is specified, then one of the conflicting
commands can be chosen from $M^o$. By Proposition \ref{prop:indep} if
$\sigma\in A$ is not in conflict with any command in $B$, then $\sigma$ is
in every merger of $A$ and $B$; in particular, it is in $M^o$ by the third
invariant. Thus, if no conflicting commands remain in $M^o$, but there is a
conflict, then $M^o$ would be a proper subset of the merger returned by the
algorithm, which is absurd.
\end{proof}

\begin{proof}[Complexity of Algorithm \ref{alg:2}]
Both the \emph{Initialize} and \emph{Finish} steps can be done in time
and space that is linear in the input size.
Updating the sets $A$ and $B$ in the \emph{Iteration} step discards some of their
elements which will never be used again. It means the total time
used by the updates is bounded by the number of elements in the original
input set $A^o\cup B^o$. Finding conflicting pairs in constant time,
however, requires building a structure which stores the bipartite conflict
graph. This requires time and space which is proportional to the number of
nodes (commands in $A^o$ and $B^o$) plus the number of conflicting pairs.
It means that Algorithm \ref{alg:2} can be implemented with both time and
space complexity that is linear in the input size plus the total number of
conflicts. While the latter number can be expected to be comparable to the
input size, in the worst case it can be quadratic, meaning that this
implementation guarantees at most quadratic time and space complexity.
\end{proof}

We conjecture that a more efficient implementation of Algorithm \ref{alg:2}
can be created using the structural properties of the conflict graph.
Some of those properties are discussed in the next Section.

\subsection{Structural properties of the conflict graph}\label{subsec:sp}

We call a conflict between two commands \emph{structural} if the out
types of the commands are different, and a \emph{content conflict}
otherwise. An example for a content conflict is when both replicas replace
the same directory with different file contents. The best resolution of a
content conflict cannot necessarily be achieved by the above winner/loser
paradigm, as it uses filesystem-level information only. Fortunately, content
conflicts are represented in the conflict graph as isolated edges. It means
that all of these conflicts must (and can) be resolved independently of all
the other conflicts. We deem it good practice to resolve all content
conflicts first.

If $(\sigma,\tau)$ is a conflicting pair and $\sigma'\in A$ is on a node
which is above the node of $\sigma$, then $(\sigma',\tau)$ is also a
conflicting pair. Proposition \ref{prop:orel} implies the same when $\sigma
\orel \sigma'$. Consequently all elements of a \emph{constructor} cluster
(see Section \ref{subsec:csets}) of $A$ are in conflict with exactly the
same elements of $B$, and so the whole constructor cluster can be replaced
by a single graph node, thus reducing the graph size. 
\ref{fig:conflict-graph} shows the typical structure of conflict graphs.
Elements of the destructor chain $\sigma_5 \orel \sigma_4 \orel \cdots \orel
\sigma_1$ are connected to more and more conflicting commands (this is so as
the corresponding nodes go upwards). Choosing $\tau_5$ as a winner
automatically resolves all conflicts by wiping out all $\sigma_i$.

Automatic conflict-based synchronizers typically give precedence to
constructors in conflicts between a constructor and a destructor command
\cite{syncpal-thesis,TSR15}, arguing that it is easier to remove some
content again than recreating it. While this is good guidance most of the
time, one has to be aware that in the current framework, replacing a
directory node with some file content is a \emph{destructor}. Confronted
with the conflicting constructor command that creates a file under the same
directory, marking this latter command as the winner is not self-evident.

\section{Conclusions}\label{sec:conclusion}

We have presented a complete theoretical investigation of filesystem
synchronization. Our definition of a filesystem is arguably simplistic, but
it captures the important features of real-word implementations. The changes
between the original and the updated filesystems are encoded using a
sequence of virtual filesystem commands. The synchronizer receives two
sequences corresponding to the two replicas to be synchronized, and produces
a third sequence which describes the synchronized, or merged state.
Definition \ref{def:merger} gives an intuitively correct and convincing
description of when the resulting sequence leads to a meaningful merged
filesystem. In Section \ref{sec:reconciler} we proved an operational
characterization of the possible merged filesystems. The main result of
Section \ref{sec:conflict} is a conflict-based non-deterministic
synchronization algorithm which creates exactly those merger sequences which
are allowed by Definition \ref{def:merger}. The simplicity of this algorithm
and the complexity of proving its correctness was a surprise to us. Its
success underlines the power of the algebraic framework of file
synchronization developed in \cite{Csi16}.

Multiple questions and possible extensions of the current work remain which
are the subject of future research. Some of them are discussed in the next
subsections.

\subsection{Synchronizing more than two replicas}

The first natural question is to extend the synchronization algorithm from
two to three or more replicas. An encouraging observation is that if several
canonical sets are pairwise refluent (for any pair there is a filesystem on
which both of them work), then they are \emph{jointly refluent}, meaning
that there is a single filesystem on which all of them work. The natural
extension of Definition \ref{def:merger} for three canonical sets $A$, $B$
and $C$ is to define their merger as a maximal canonical set $M\subseteq
A\cup B\cup C$. This definition has the additional advantage that it is
symmetric in the filesystems, and does not give precedence to any of them
over the others. We can prove that such a merger also satisfies an
operational characterization similar to the one stated in Theorem
\ref{thm:merge}. It seems that such a 3-way merger can be generated first by
merging $A$ and $B$ to $M$, and then merging $M$ and $C$. Is this claim
true?

Another question is whether the general task of finding a 3-way merger can
be reduced to the case when $A$, $B$ and $C$ are pairwise disjoint. Finally,
at least in the case when the sets $A$, $B$, $C$ are pairwise disjoint,
whether the following variant of the Iteration step of Algorithm \ref{alg:2}
generates a 3-way merger, and if yes, whether it generates all
possibilities:

\bgroup\smallskip
\hangafter1\hangindent\parindent\parindent=0pt
\textit{Iterate:} Pick a winner command
$\sigma$ from $A\cup B\cup C$, and discard those commands from the other two
sets which are in conflict with $\sigma$.
\par\egroup

\subsection{Efficiency of the algorithms}

Algorithm \ref{alg:1} can be implemented with a runtime that increases
linearly with the size of its input, so it is efficient. The trivial running
time estimation of Algorithm \ref{alg:2} is proportional to the product of
the sizes of the command sets $A$ and $B$ -- assuming that the ``above''
relation in the node structure $\N$ can be checked in constant time, as
generating the conflict graph takes that much time. Section \ref{subsec:sp}
discusses some possible improvements by exploiting the structure of the
conflict graph. As to whether the running time of the algorithm can be
improved significantly, or it is substantially quadratic, our conjecture is
that the algorithm can be made sub-quadratic.

\subsection{Attributes}\label{subsec:attributes}

Attributes typically store metainformation about a node, such as different
timestamps (e.g. creation date, last modification or last access) and
permissions (e.g. read and write access), and whether the metainformation
refers to the node only or to the whole subtree below it. Attributes are set
and modified either automatically, or by special filesystem commands. In our
filesystem model attributes can be best modeled by storing them as node
content. For file nodes it means that the attributes are added to the actual
file content, which means that any change to the attributes can be handled
as a change in content. Therefore, file attributes pose no special problems
and can be integrated seamlessly into our model. Handling directory
attributes, however, is a challenging open problem. A general algebraic
framework of filesystems allowing content in directory (and even in empty)
nodes was developed in \cite{CC22} at the expense of more complicated
notions and theorems. Unfortunately, with this extension many of the
Propositions in Sections \ref{sec:reconciler} and \ref{sec:conflict} on
which the correctness of our method relies do not remain true. To illustrate
the problem, suppose $A$ changes some attribute from ``private'' to
``public'' at some directory node $n$, and creates a file under $n$; while
$B$, under the impression that the directory is still private, creates
another file under it. Then the file creation command from $B$ cannot be
unconditionally moved to the merging sequence.

One possible solution is to consider and resolve all attribute conflicts
(including when $A$ and $B$ modify attributes at the same node differently)
as the first stage of the synchronization process, and then proceed
according to the algorithms described in this paper. This approach, however,
is not necessarily compatible with our declarative Definition
\ref{def:merger} of what a synchronized state is. It is not even clear
whether all possible mergers satisfying this definition are intuitively
correct or not if directory attributes are taken into account.

\subsection{Links}\label{subsec:links}

From the user's perspective, a \emph{link} between the nodes $n$ and $n'$ is
a promise, or a commitment, that the filesystem at and below $n$ is exactly
the same as at and below $n'$. Executing a command below $n$ automatically
executes the same command on the corresponding node below $n'$. A link
system must be \emph{loopless} meaning that any node has only finitely many
other nodes \emph{equivalent} to it. In particular, $n$ and $n'$ must be
uncomparable nodes.

If the original filesystem $\FS$ contains links, but the users do not have
tools or permissions to manipulate them, the following synchronization
method works. The update detector, knowing the link structure, unfolds it,
and replaces each user command by the collection of all ``hidden'' or
``implicit'' commands on the equivalent nodes. After creating the canonical
command set $A$, it marks the commands which are on equivalent nodes. For
the merger $M$ to be applicable to $\FS$ it must also contain, with each
command in $M$, all of its equivalents. This can be achieved by the
following modification to the \emph{Iterate} part of Algorithm \ref{alg:2}:
if $\sigma$ is chosen as a winner, then all commands equivalent to it will
be winners as well.

It is an open problem to introduce filesystem commands which manipulate the
link structure and can be incorporated into our synchronization process.

\vspace{6pt} 


\funding{The research of the second author (L.Cs) has been supported by the
GACR project number 19-04579S and by the Lendulet Program of the HAS, which
is thankfully acknowledged.}



\abbreviations{Abbreviations}{
The following abbreviations are used in this manuscript:\\

\noindent 
\begin{tabular}{@{}ll}
CRDT & Conflict-free Replicated Data Type\\
CSCW & Computer Supported Collaborative Work\\
OT & Operational Transformation\\
$\E$, $\F$, $\D$ & empty, file, and directory content\\
$\FS$, $\FSa$ & filesystem\\
\end{tabular}
}

\begin{adjustwidth}{-\extralength}{0cm}
\reftitle{References}

\end{adjustwidth}
\end{document}